%% file: mdpr.tex
\pdfoutput=1
\documentclass[sigconf]{acmart}

\usepackage{enumitem}
\usepackage{booktabs}
\usepackage{xspace}
\usepackage{multirow}
\usepackage{subcaption}
\usepackage{caption}
\usepackage{stackengine}
\def\delequal{\mathrel{\ensurestackMath{\stackon[1pt]{=}{\scriptstyle\Delta}}}}
\usepackage{graphicx}
\usepackage{xcolor,colortbl}

\newcommand\ignore[1]{}

\newcommand\mrtydi{Mr.~T{\small Y}D{\small I}\xspace}
\newcommand\tydi{T{\small Y}D{\small I}\xspace}
\newcommand\mbert{mBERT\xspace}

\setcopyright{none}
\renewcommand\footnotetextcopyrightpermission[1]{}

\begin{document}

\title[]{Towards Best Practices for Training\\ Multilingual Dense Retrieval Models}

\author{Xinyu Zhang$^*$, 
Kelechi Ogueji$^*$, Xueguang Ma, Jimmy Lin}
\thanks{$^*$ Equal Contribution}

\affiliation{\vspace{0.1cm}
David R. Cheriton School of Computer Science \\
University of Waterloo, Ontario \country{Canada}
}

\renewcommand{\shortauthors}{}
\pagestyle{empty}

\begin{abstract}
Dense retrieval models using a transformer-based bi-encoder design have emerged as an active area of research.
In this work, we focus on the task of monolingual retrieval in a variety of typologically diverse languages using one such design.
Although recent work with multilingual transformers demonstrates that they exhibit strong cross-lingual generalization capabilities, there remain many open research questions, which we tackle here.
Our study is organized as a ``best practices'' guide for training multilingual dense retrieval models, broken down into three main scenarios:\ where a multilingual transformer is available, but relevance judgments are not available in the language of interest; where both models and training data are available; and, where training data are available not but models.
In considering these scenarios, we gain a better understanding of the role of multi-stage fine-tuning, the strength of cross-lingual transfer under various conditions, the usefulness of out-of-language data, and the advantages of multilingual vs.\ monolingual transformers.
Our recommendations offer a guide for practitioners building search applications, particularly for low-resource languages, and while our work leaves open a number of research questions, we provide a solid foundation for future work.
\end{abstract}

\maketitle

\input{results/colors}

\section{Introduction}

Retrieval based on dense vector representations derived from pretrained transformers such as BERT~\cite{devlin-etal-2019-bert} represents an active area of research.
By adopting a so-called bi-encoder (or dual-encoder) architecture, document representations can be computed offline and document ranking can be recast as a nearest neighbor search problem given the query vector representation, for which existing open-source toolkits such as Faiss~\cite{Johnson_etal_2021} or nmslib~\cite{Boytsov_etal_CIKM2016} provide efficient and scalable solutions.
A bi-encoder design provides an attractive alternative to the so-called cross-encoder design, whereby queries and documents are concatenated and fed into a transformer directly.
Cross-encoders are only practical in a reranking setup, processing candidates generated by a first-stage retrieval, whereas bi-encoders directly support single-stage retrieval (but of course, can be further reranked by cross-encoders if desired).

Our work explores dense retrieval methods based on bi-encoders in a multilingual context.
To be clear, we are concerned with monolingual retrieval, particularly in non-English languages, where both queries and documents are in the {\it same} language $\mathcal{L}$ (which we call the target language).
While many of our techniques are also applicable to the cross-lingual case, where queries and documents are in {\it different} languages, we do not explicitly consider this application.
Recent work has shown that retrieval models (bi-encoders and cross-encoders) based on multilingual transformers exhibit strong cross-lingual generalization capabilities.
That is, we can train the model using relevance judgments in one language and apply inference in another language for ranking~\cite{MacAvaney2020TeachingAN,shi-etal-2021-cross,asai-2021-one,mrtydi,Bonifacio:2108.13897:2022}.
It appears that models are able to ``transfer'' relevance matching capabilities across languages.
This is an exciting finding because it becomes possible to leverage relevance judgments in one language (for example, a high-resource language such as English) to build retrieval models for low resource languages (e.g., Telugu).

Of course, the reality is quite complex (see additional discussions in  Section~\ref{section:background}) and there remain many open research questions.
Suppose we are interested in building a dense retrieval model for target language $\mathcal{L}$.
What should we do if we don't have any training data in $\mathcal{L}$?
Should we rely entirely on cross-lingual transfer from another dataset (for example, MS MARCO in English)?
What if we {\it do} have (limited) data in $\mathcal{L}$?
Is a large dataset such as MS MARCO in English still helpful?
Transformers can be fine-tuned in different ``stages''~\cite{Garg_etal_AAAI2020,Xie_etal_WWW2020,ZhangXinyu_etal_ECIR2021}, so it is certainly possible to exploit datasets in different languages using a multilingual transformer backbone.
Should we?
Does the answer change if the datasets draw from linguistically unrelated families?
Or written in entirely different scripts?
And to consider an entirely different approach, should we prefer monolingual transformers if one is available?

The goal of our work is to answer these myriad questions, and to begin to develop a set of best practices for training multilingual dense retrieval models.
Our exploration is organized into three main scenarios, as follows:

\begin{enumerate}[leftmargin=*]

\item {\bf Have model, no data} --- scenario (1): The target language $\mathcal{L}$ is covered by a multilingual transformer, but unfortunately we don't have any training data.

\item {\bf Have model and data} --- scenario (2): The target language $\mathcal{L}$ is covered by a multilingual transformer and we have training data. This breaks down into two sub-cases, one where we actually have data in language $\mathcal{L}$ and the other where we have data in some other language.
We also consider the case where a {\it monolingual} transformer is available.

\item {\bf Have data, no model} --- scenario (3): No pretrained transformer is available for the target language $\mathcal{L}$, but we do have data in the target language.

\end{enumerate}

\noindent In this work, we build on Dense Passage Retriever (DPR)~\cite{dpr}, one of the earliest and most foundational dense retrieval models in the literature.
We view our contributions as follows:

\begin{enumerate}[leftmargin=*]

\item We provide concrete guidance and recommended best practices, supported by empirical results, on how to train multilingual DPR models for monolingual retrieval.

\item In the course of our study, we answer a number of scientific questions about multilingual dense retrieval models that until now have not been clearly addressed.
For example, we find that ``pre--fine-tuning'' with the MS MARCO passage dataset (in English) rarely hurts effectiveness, even if the target language is unrelated to English.

\item Many of our results are surprising, such as the fact that multilingual BERT, fine-tuned on {\it Thai}, is relatively effective for retrieval in {\it Arabic}, and that monolingual {\it English} BERT, fine-tuned on {\it Russian} relevance judgments, is effective for retrieval in {\it Russian}.
We begin to untangle {\it why} in terms of cross-lingual anchors and token overlap across corpora in different languages.

\end{enumerate}

\noindent In summary, we provide practitioners with a ``how to'' guide for building multilingual search applications and researchers with a solid foundation for future work in multilingual dense retrieval.

\section{Background and Related Work}
\label{section:background}

{\bf mDPR.}
We base our retrieval models on DPR~\citep{dpr}, a foundational and one of the earliest dense retrieval models.
Given a question $q$ and a passage $p$, DPR generates representations, $E_Q(q)$ and $E_P(p)$, independently using question and passage encoders based on BERT by taking the representation of the \textsc{[CLS]} token in the final layer.
Similarity between a query and a passage is measured by the inner product of their representations:
    \begin{align*}
        s_{q, p} \delequal \textrm{sim}(q, p) = E_Q(q)^TE_P(p)
    \end{align*}
which is optimized according to the NCE loss:
\begin{align*}
    L(q_i, p_i^+, p_{i,1}^-, ..., p_{i,n}^-) = 
    - \log( \frac
    { \exp(s_{q_i, p_i^+}) }
    { \exp(s_{q_i, p_i^+}) + \sum^n_{j=1}\exp(s_{q_i, p_{i,j}^-})}
    )
\end{align*}
where $q_i$ is the question representation, 
$p_i^+$ is the representation of a corresponding positive passage,
and $p_j^-$ is the representation of a negative passage.
The similarity scores $s_{q, p}$ are used to generate a ranking for query $q$, using a nearest neighbor vector search library (in our experiments, Faiss).
Although we selected DPR for our experiments, any alternative model such as ANCE~\cite{xiong21ance}, TCT-ColBERT~\cite{tct_colbert}, or TAS-B~\cite{tasb} can be substituted; the exact choice is unlikely to affect our overall findings.

DPR is originally initialized with English BERT~\citep{dpr}. 
Following previous work~\cite{asai-2021-one,mrtydi}, for retrieval in multiple languages and to exploit cross-lingual transfer, we instead initialized the model with multilingual BERT (mBERT), keeping all other aspects of training identical.
We call this ``mDPR''.

\begin{table}[!t]
\centering
\resizebox{1.0\columnwidth}{!}{
    \begin{tabular}{ll|rr|rr|rr|r}
    \toprule
    \multirow{2}{*}{} &
        & \multicolumn{2}{c}{\textbf{Train}} & \multicolumn{2}{c}{\textbf{Dev}} & \multicolumn{2}{c}{\textbf{Test}} & 
    \multirow{2}{*}{\textbf{Corpus Size}} \\
        & & \textbf{\# Q} & \textbf{\# J} & \textbf{\# Q} & \textbf{\# J} & \textbf{\# Q} & \textbf{\# J} & \\
    \midrule
    Arabic & (Ar) 
        & 12,377 & 12,377 & 3,115 & 3,115 & 1,081 & 1,257 & 2,106,586 \\
    Bengali & (Bn) 
        & 1,713 & 1,719 & 440 & 443 & 111 & 130 & 304,059 \\
    English $^{\ddagger}$  & (En) 
        & 3,547 & 3,547 & 878 & 878 & 744 & 935 & 32,907,100 \\
    Finnish $^{\ddagger}$ & (Fi)
         & 6,561 & 6,561 & 1,738 & 1,738 & 1,254 & 1,451 & 1,908,757 \\
    Indonesian $^{\ddagger}$  & (Id)
        & 4,902 & 4,902 & 1,224 & 1,224 & 829 & 961 & 1,469,399 \\
    Japanese & (Ja) 
        & 3,697 & 3,697 & 928 & 928 & 720 & 923 & 7,000,027 \\
    Korean & (Ko) 
        & 1,295 & 1,317 & 303 & 307 & 421 & 492 & 1,496,126 \\
    Russian & (Ru) 
        & 5,366 & 5,366 & 1,375 & 1,375 & 995 & 1,168 & 9,597,504 \\
    Swahili $^{\ddagger}$  & (Sw) 
        & 2,072 & 2,401 & 526 & 623 & 670 & 743 & 136,689 \\
    Telugu & (Te) 
        & 3,880 & 3,880 & 983 & 983 & 646 & 664 & 548,224 \\
    Thai & (Th)
        & 3,319 & 3,360 & 807 & 817 & 1,190 & 1,368 & 568,855 \\
    \midrule
    Total &
        & 48,729 & 49,127 & 12,317 & 12,431 & 8,661 & 10,092 & 58,043,326 \\
    \bottomrule
    \end{tabular}
}
\vspace{0.1cm}
\caption{Statistics for \mrtydi:\ number of questions (\#~Q), judgments (\#~J), and number of passages (Corpus Size) in each language. $^{\ddagger}$ indicates Latin-script languages.}
\vspace{-0.5cm}
\label{tab:mr-tydi-stats}
\end{table}

\smallskip \noindent
{\bf \mrtydi.}
All experiments and recommendations in this work are based on \mrtydi~\cite{mrtydi}, a multilingual retrieval benchmark based on the \tydi~\cite{tydi} dataset (originally developed for question answering).
\mrtydi covers 11 languages from different language families that are typologically diverse.
Among the languages, English, Finnish, Indonesian, and Swahili are written in Latin script, whereas no other language pair shares the same script.
The corpus for each language is drawn from Wikipedia, and the query and judgements are prepared by native speakers of that language.
Table~\ref{tab:mr-tydi-stats} presents statistics of the \mrtydi dataset, copied from the original paper.
To our knowledge, this is the only available dataset that meets our needs (i.e., monolingual retrieval); other datasets, for example, XOR-\tydi~\cite{Asai:2010.11856:2021} and CLIRMatrix~\cite{sun-duh-2020-clirmatrix} focus on cross-lingual retrieval.

\smallskip \noindent
{\bf Multilingual Retrieval.}
There is a large body of literature on cross-lingual retrieval dating back decades.
This is {\it not} our focus, and a comprehensive review is not possible in this limited space.
Nevertheless, we refer readers to~\citet{Nie_2010} for a survey of pre-neural techniques and~\citet{clir-survey} for a more recent review.

While \mbert provides a convenient starting point for exploring cross-lingual transfer in information retrieval,
the study of this phenomenon predates the emergence of \mbert. 
\citet{zhang-etal-2019-improving-low} found that aligning the query and documents of a source and target language could aid
transfer ability in the Cross-Lingual Word Embedding (CLWE) setting.
More recently, \citet{litschko-2021-evaluating-multilingual} compared the effectiveness of CLWE models and \mbert under an unsupervised scenario in the cross-lingual setting.

Since the emergence of multilingual pretrained language models, researchers have been exploiting cross-lingual transfer to improve monolingual retrieval.
\citet{MacAvaney2020TeachingAN} and \citet{Shi2020CrossLingualTO} both explored the effectiveness of zero-shot transfer, where mBERT is fine-tuned on a source language and directly applied to a target language.
However, these papers exploited cross-encoders, as opposed to the bi-encoders that we explore here.
While cross-encoders can be more effective, they are often much slower since they operate in a retrieve-and-rerank setup.
In contrast, the bi-encoder design we explore is amenable to single-stage retrieval with nearest neighbor vector search frameworks such as Faiss.
\citet{asai-2021-one} also made use of mBERT for retrieval in a many-to-many scenario.
They train a QA model that can answer a query in any language by retrieving evidence from a multilingual collection and generating an answer in the query language.
This differs from our setting where we are strictly focused on retrieving relevant documents from a monolingual collection given a query in the same language.

\smallskip \noindent
{\bf \mbert's Cross-lingual Capabilities.}
There has been much recent work on trying to understand the impressive cross-lingual abilities of \mbert, with many conflicting results.
Among the earliest demonstration was the work of \citet{wu-dredze-2019-beto}, who evaluated mBERT in a zero-shot cross-lingual transfer scenario on 5 NLP tasks across 39 languages and found that mBERT achieves impressive effectiveness.
They argued that shared subwords aid cross-lingual transfer, based on a strong correlation between the percentage of overlapping subwords and transfer effectiveness.
\citet{pires-etal-2019-multilingual} also reported strong cross-lingual transfer abilities using mBERT, highlighting the model's effectiveness on transferring across language scripts and code-switched text.
Regarding the reasons for the model's ability to generalize across languages, they hypothesized that common wordpieces such as numbers and URLs are responsible, since these are the same across languages and thus share representations.
\citet{dufter-schutze-2020-identifying} echoed this finding, claiming that shared position embeddings and shared special tokens are necessary ingredients for cross-lingual capabilities.

However, there have been contradictory results as well.
\citet{wang2019cross} explored the contribution of different components in \mbert to its cross-lingual capabilities and concluded that token overlap between languages does not play a significant role.
They argued that other parts of the architecture, such as the depth of the network, are more critical for creating a multilingual model.
\citet{artetxe-etal-2020-cross} also showed that a shared vocabulary or joint pretraining are not needed to obtain cross-lingual transfer.
Using this knowledge, they were able to induce multilinguality from monolingual language models.
\citet{conneau-etal-2020-emerging} also echoed the finding about the insignificance of token overlaps in cross-lingual transfer.
They also showed that shared softmax and domain similarity play similarly minor roles. 
Instead, they found that parameter sharing has the biggest impact in learning cross-lingual representation.

While these papers have provided interesting but inconclusive and contradictory findings, they have all been focused on NLP tasks.
To our knowledge, we are the first to systematically study and categorize the cross-lingual transfer abilities of mBERT and to attempt an explanation for the (often counter-intuitive) observed behaviors, specifically for a retrieval task.

\section{Experimental Design}

\subsection{Data Preparation}
\label{sec:data-prep}

The experiments in this work are all based on \mrtydi.
We use the version 1.1 dataset released on HuggingFace.\footnote{\url{https://huggingface.co/datasets/castorini/mr-tydi}}
In the training set used to train our dense retrieval models, the positive examples are from the labeled positive passages provided by \mrtydi, and the negative examples are prepared from the top-30 results retrieved from the tuned BM25 baseline described in \citet{mrtydi}; using BM25 to generate negatives is standard practice~\cite{dpr}.

In our experiments, we train models with data from a single language and with data combined from multiple languages.
In the multiple languages case, data in the same batch are always drawn from the same language.
Pilot experiments indicated that this practice is helpful because it prevents in-batch negatives from degenerating into a language detection task, where the model tries to distinguish positive passages (in one language) from negative passages (in another language).

For experiments in Section~\ref{sec:reuslts-case2}, where we fine-tune in language $\mathcal{K}$ and test on $\mathcal{L}$ in a large matrix experiment, we wanted to alleviate the effects of training data size, since different languages have different numbers of relevance judgments.
However, based on pilot experiments, simply down-sampling all datasets to the smallest language yielded poor results.
Instead, we randomly sampled 3300 training queries from each language, which is close to the size of training queries in Thai.
For Bengali, Korean, and Swahili, which have fewer judgments, we retained all available data.
Pilot experiments confirmed that sampling fewer queries would result in under-fitting for most of the languages.
It is worth emphasizing that down-sampling was applied {\it only} for the matrix experiments in Section~\ref{sec:reuslts-case2}; the remaining experiments used all available data.

\begin{table}[t]
    \centering
    \resizebox{0.8\columnwidth}{!}{
        \begin{tabular}{c|c}
        \toprule
            Language & HuggingFace Model \\
        \midrule
            mBERT & \texttt{bert-base-multilingual-cased} \\
            AfriBERTa & \texttt{castorini/afriberta\_large} \\
            Arabic &  \texttt{asafaya/bert-base-arabic} \\
            English &  \texttt{bert-base-uncased} \\
            Finnish &  \texttt{TurkuNLP/bert-base-finnish-cased-v1} \\
            Indonesian &  \texttt{cahya/bert-base-indonesian-522M} \\
            Korean &  \texttt{kykim/bert-kor-base} \\
        \bottomrule
        \end{tabular}
    }
    \vspace{0.1cm}
    \caption{HuggingFace models used in our experiments.
    }
    \label{tab:plm-models}
\vspace{-0.5cm}
\end{table}

\input{results/main-results-table}

\subsection{Models}
\label{sec:config-model}

Our models are trained using Tevatron,\footnote{https://github.com/texttron/tevatron} a more efficient implementation of the original DPR code open-sourced by~\citet{dpr}.
We trained different dense retrieval models with shared parameters between the query and passage encoders, initialized with different pretrained language model checkpoints, all of which are available on HuggingFace~\cite{wolf-etal-2020-transformers} (see Table~\ref{tab:plm-models}).
Of course, our main focus is on mBERT, but as additional points of comparison, we also tried training monolingual DPR with monolingual BERT (in non-English languages) to answer the research questions posed in Section~\ref{section:mono}.
Unfortunately, not all languages have publicly available monolingual pretrained BERT models of good quality.
Furthermore, we only considered monolingual models that were trained from scratch (not initialized with existing models).
In the end, we experimented with monolingual BERT models for five languages in \mrtydi:\ Arabic, English, Finnish, Indonesian, and Korean.

In all experiments, models were trained 40 epochs with the corresponding training data (single or multiple languages) with 128 batch size.
We used the Adam optimizer, setting the learning rate to 4e-5 for experiments trained on a single language, and 1e-5 for others. 
The maximum length of queries and passages is set to 64 and 256, respectively.\footnote{
We tuned the learning rate and the number of epochs on the development set, among \{5e-6, 1e-5, 2e-5, 4e-5, 8e-5\} and \{10, 20, 30, 40, 50\}, respectively.
}
Following~\citet{mrtydi}, we report MRR@100 and Recall@100 on the test sets of each language in \mrtydi.

For experiments involving BM25 and mDPR hybrid, we followed exactly the sparse--dense hybrid approach in \citet{mrtydi}, where the final score $s_{\text{hybrid}}$ is computed as a linear combination of the BM25 score $s_{\text{sparse}}$ and the dense score $s_{\text{dense}}$, with $\alpha$ as a weighting parameter tuned on the development set.

In our analyses, we report statistically significant differences detected using paired $t$-tests ($p<0.01$).
Wary of the dangers of multiple hypothesis testing, we do not apply tests indiscriminately, but rather only to answer specific research questions.

\section{Have Model, No Data!}
\label{section:case1}

{\bf Scenario (1)}: $\mathcal{L} \in \textrm{mBERT}$, but we have no data in $\mathcal{L}$.

\smallskip \noindent {\bf Recommendation: }
pre--fine-tune mDPR with an mBERT backbone (from raw checkpoint) on MS MARCO passage (in English), and then apply retrieval directly in $\mathcal{L}$ in a zero-shot manner. $\square$

\medskip
\noindent In this section, we provide supporting experimental evidence of how we arrived at the above recommendation, primarily drawn from Table~\ref{table:main}.
Here, each row represents an experimental condition, and each column shows the effectiveness (MRR@100 on top, Recall@100 on bottom) for each language.
Rows (1) and (2) capture BM25 with default and tuned parameters, copied directly from the original \mrtydi paper~\cite{mrtydi}.
Row~(3) is also copied from that paper, representing a zero-shot baseline.
In this condition, mDPR is fine-tuned with the Natural Questions dataset and then evaluated on each of the languages directly, without the model having seen any annotated examples from \mrtydi.

Building on this, the obvious improvement is to follow the same process as row~(3), but using a different dataset.
In row~(4), we fine-tune mDPR on the MS MARCO passage ranking data, which is much larger than NQ, and then apply inference for all languages in a zero-shot manner.
In both cases, rows~(3) and (4), we refer to this process as pre--fine-tuning (or pFT for short)~\cite{ZhangXinyu_etal_ECIR2021}, to distinguish additional fine-tuning that we can further perform downstream (discussed later).

We see clearly that pFT with MS MARCO passage, row~(4), consistently beats pFT with NQ, row~(3); these differences are statistically significant across all languages and support our recommendation that if one doesn't have data in $\mathcal{L}$, pre--fine-tuning on a large existing dataset is the best course of action.
This strategy takes advantage of cross-lingual transfer effects, and is consistent with previous work.
The dataset used for pre--fine-tuning {\it does} matter, and here we find that MS MARCO passage is better than NQ.
This suggests that dataset size is a more important factor than domain match:\ Although NQ is smaller, it is ``closer'' to \mrtydi---both contain relatively well-formed questions posed against Wikipedia---whereas MS MARCO contains more noisy and (sometimes) ill-formed questions, annotated against the general web.

\input{results/lang2lang-table}

\begin{table}[t]
\resizebox{\columnwidth}{!}{%
\begin{tabular}{cccccccccccc}
\hline
 & \textbf{Ar} & \textbf{Bn} & \textbf{En} & \textbf{Fi} & \textbf{In} & \textbf{Ja} & \textbf{Ko} & \textbf{Ru} & \textbf{Sw} & \textbf{Te} & \textbf{Th} \\ \hline

Ar & \cellcolor{diff-w-ms_1_0}{0.218} & \cellcolor{diff-w-ms_1_1}{0.224} & \cellcolor{diff-w-ms_1_2}{0.135} & \cellcolor{diff-w-ms_1_3}{0.144} & \cellcolor{diff-w-ms_1_4}{0.130} & \cellcolor{diff-w-ms_1_5}{0.119} & \cellcolor{diff-w-ms_1_6}{0.095} & \cellcolor{diff-w-ms_1_7}{0.172} & \cellcolor{diff-w-ms_1_8}{0.056} & \cellcolor{diff-w-ms_1_9}{0.251} & \cellcolor{diff-w-ms_1_10}{0.097}\\
Bn & \cellcolor{diff-w-ms_2_0}{0.105} & \cellcolor{diff-w-ms_2_1}{0.261} & \cellcolor{diff-w-ms_2_2}{0.117} & \cellcolor{diff-w-ms_2_3}{0.121} & \cellcolor{diff-w-ms_2_4}{0.090} & \cellcolor{diff-w-ms_2_5}{0.118} & \cellcolor{diff-w-ms_2_6}{0.104} & \cellcolor{diff-w-ms_2_7}{0.098} & \cellcolor{diff-w-ms_2_8}{0.069} & \cellcolor{diff-w-ms_2_9}{0.281} & \cellcolor{diff-w-ms_2_10}{0.134}\\
En & \cellcolor{diff-w-ms_3_0}{0.024} & \cellcolor{diff-w-ms_3_1}{0.029} & \cellcolor{diff-w-ms_3_2}{0.164} & \cellcolor{diff-w-ms_3_3}{-0.014} & \cellcolor{diff-w-ms_3_4}{-0.059} & \cellcolor{diff-w-ms_3_5}{0.055} & \cellcolor{diff-w-ms_3_6}{0.047} & \cellcolor{diff-w-ms_3_7}{0.039} & \cellcolor{diff-w-ms_3_8}{-0.067} & \cellcolor{diff-w-ms_3_9}{-0.080} & \cellcolor{diff-w-ms_3_10}{-0.038}\\
Fi & \cellcolor{diff-w-ms_4_0}{0.098} & \cellcolor{diff-w-ms_4_1}{0.118} & \cellcolor{diff-w-ms_4_2}{0.156} & \cellcolor{diff-w-ms_4_3}{0.218} & \cellcolor{diff-w-ms_4_4}{0.117} & \cellcolor{diff-w-ms_4_5}{0.082} & \cellcolor{diff-w-ms_4_6}{0.112} & \cellcolor{diff-w-ms_4_7}{0.140} & \cellcolor{diff-w-ms_4_8}{0.072} & \cellcolor{diff-w-ms_4_9}{0.100} & \cellcolor{diff-w-ms_4_10}{0.073}\\
In & \cellcolor{diff-w-ms_5_0}{0.107} & \cellcolor{diff-w-ms_5_1}{0.155} & \cellcolor{diff-w-ms_5_2}{0.116} & \cellcolor{diff-w-ms_5_3}{0.135} & \cellcolor{diff-w-ms_5_4}{0.183} & \cellcolor{diff-w-ms_5_5}{0.071} & \cellcolor{diff-w-ms_5_6}{0.099} & \cellcolor{diff-w-ms_5_7}{0.106} & \cellcolor{diff-w-ms_5_8}{0.061} & \cellcolor{diff-w-ms_5_9}{0.133} & \cellcolor{diff-w-ms_5_10}{0.067}\\
Ja & \cellcolor{diff-w-ms_6_0}{0.087} & \cellcolor{diff-w-ms_6_1}{0.116} & \cellcolor{diff-w-ms_6_2}{0.059} & \cellcolor{diff-w-ms_6_3}{0.061} & \cellcolor{diff-w-ms_6_4}{0.027} & \cellcolor{diff-w-ms_6_5}{0.169} & \cellcolor{diff-w-ms_6_6}{0.099} & \cellcolor{diff-w-ms_6_7}{0.116} & \cellcolor{diff-w-ms_6_8}{0.016} & \cellcolor{diff-w-ms_6_9}{0.066} & \cellcolor{diff-w-ms_6_10}{0.092}\\
Ko & \cellcolor{diff-w-ms_7_0}{0.071} & \cellcolor{diff-w-ms_7_1}{0.067} & \cellcolor{diff-w-ms_7_2}{0.115} & \cellcolor{diff-w-ms_7_3}{0.073} & \cellcolor{diff-w-ms_7_4}{0.050} & \cellcolor{diff-w-ms_7_5}{0.112} & \cellcolor{diff-w-ms_7_6}{0.156} & \cellcolor{diff-w-ms_7_7}{0.162} & \cellcolor{diff-w-ms_7_8}{0.044} & \cellcolor{diff-w-ms_7_9}{0.041} & \cellcolor{diff-w-ms_7_10}{0.054}\\
Ru & \cellcolor{diff-w-ms_8_0}{0.125} & \cellcolor{diff-w-ms_8_1}{0.138} & \cellcolor{diff-w-ms_8_2}{0.133} & \cellcolor{diff-w-ms_8_3}{0.131} & \cellcolor{diff-w-ms_8_4}{0.110} & \cellcolor{diff-w-ms_8_5}{0.114} & \cellcolor{diff-w-ms_8_6}{0.103} & \cellcolor{diff-w-ms_8_7}{0.162} & \cellcolor{diff-w-ms_8_8}{0.052} & \cellcolor{diff-w-ms_8_9}{0.147} & \cellcolor{diff-w-ms_8_10}{0.048}\\
Sw & \cellcolor{diff-w-ms_9_0}{0.054} & \cellcolor{diff-w-ms_9_1}{0.086} & \cellcolor{diff-w-ms_9_2}{0.073} & \cellcolor{diff-w-ms_9_3}{0.103} & \cellcolor{diff-w-ms_9_4}{0.062} & \cellcolor{diff-w-ms_9_5}{0.051} & \cellcolor{diff-w-ms_9_6}{0.056} & \cellcolor{diff-w-ms_9_7}{0.092} & \cellcolor{diff-w-ms_9_8}{0.266} & \cellcolor{diff-w-ms_9_9}{0.131} & \cellcolor{diff-w-ms_9_10}{0.048}\\
Te & \cellcolor{diff-w-ms_10_0}{0.131} & \cellcolor{diff-w-ms_10_1}{0.200} & \cellcolor{diff-w-ms_10_2}{0.110} & \cellcolor{diff-w-ms_10_3}{0.084} & \cellcolor{diff-w-ms_10_4}{0.055} & \cellcolor{diff-w-ms_10_5}{0.136} & \cellcolor{diff-w-ms_10_6}{0.116} & \cellcolor{diff-w-ms_10_7}{0.143} & \cellcolor{diff-w-ms_10_8}{0.063} & \cellcolor{diff-w-ms_10_9}{0.505} & \cellcolor{diff-w-ms_10_10}{0.153}\\
Th & \cellcolor{diff-w-ms_11_0}{0.143} & \cellcolor{diff-w-ms_11_1}{0.147} & \cellcolor{diff-w-ms_11_2}{0.131} & \cellcolor{diff-w-ms_11_3}{0.104} & \cellcolor{diff-w-ms_11_4}{0.112} & \cellcolor{diff-w-ms_11_5}{0.113} & \cellcolor{diff-w-ms_11_6}{0.114} & \cellcolor{diff-w-ms_11_7}{0.143} & \cellcolor{diff-w-ms_11_8}{0.059} & \cellcolor{diff-w-ms_11_9}{0.242} & \cellcolor{diff-w-ms_11_10}{0.329}\\

\bottomrule
\end{tabular}
}
\vspace{0.1cm}
\caption{
Element-wise difference between MS pFT + FT vs.\ MS pFT (zero-shot), answering the question: ``Should we always fine-tune?'' Answer is  {\it yes}, with the exception that further fine-tuning on English data for retrieval in non-English languages might result in overfitting.
Positive values are highlighted in \textcolor{blue}{blue} and negative values in \textcolor{orange}{\textbf{orange}}, with saturation proportional to magnitude.
}
\label{table:delta-to-ms0shot}
\vspace{-0.3cm}
\end{table}

\begin{table}[t]
\resizebox{\columnwidth}{!}{%
\begin{tabular}{cccccccccccc}
\hline
 & \textbf{Ar} & \textbf{Bn} & \textbf{En} & \textbf{Fi} & \textbf{In} & \textbf{Ja} & \textbf{Ko} & \textbf{Ru} & \textbf{Sw} & \textbf{Te} & \textbf{Th} \\ \hline
Ar & \cellcolor{diff-w-each-lang_1_0}{0.033} & \cellcolor{diff-w-each-lang_1_1}{0.047} & \cellcolor{diff-w-each-lang_1_2}{0.077} & \cellcolor{diff-w-each-lang_1_3}{-0.005} & \cellcolor{diff-w-each-lang_1_4}{0.025} & \cellcolor{diff-w-each-lang_1_5}{0.065} & \cellcolor{diff-w-each-lang_1_6}{0.008} & \cellcolor{diff-w-each-lang_1_7}{0.058} & \cellcolor{diff-w-each-lang_1_8}{-0.035} & \cellcolor{diff-w-each-lang_1_9}{0.107} & \cellcolor{diff-w-each-lang_1_10}{-0.007}\\
Bn & \cellcolor{diff-w-each-lang_2_0}{0.101} & \cellcolor{diff-w-each-lang_2_1}{0.006} & \cellcolor{diff-w-each-lang_2_2}{0.123} & \cellcolor{diff-w-each-lang_2_3}{0.055} & \cellcolor{diff-w-each-lang_2_4}{0.085} & \cellcolor{diff-w-each-lang_2_5}{0.118} & \cellcolor{diff-w-each-lang_2_6}{0.075} & \cellcolor{diff-w-each-lang_2_7}{0.110} & \cellcolor{diff-w-each-lang_2_8}{0.006} & \cellcolor{diff-w-each-lang_2_9}{0.112} & \cellcolor{diff-w-each-lang_2_10}{0.076}\\
En & \cellcolor{diff-w-each-lang_3_0}{0.029} & \cellcolor{diff-w-each-lang_3_1}{0.023} & \cellcolor{diff-w-each-lang_3_2}{0.044} & \cellcolor{diff-w-each-lang_3_3}{-0.089} & \cellcolor{diff-w-each-lang_3_4}{-0.083} & \cellcolor{diff-w-each-lang_3_5}{0.072} & \cellcolor{diff-w-each-lang_3_6}{0.030} & \cellcolor{diff-w-each-lang_3_7}{0.013} & \cellcolor{diff-w-each-lang_3_8}{-0.095} & \cellcolor{diff-w-each-lang_3_9}{-0.006} & \cellcolor{diff-w-each-lang_3_10}{0.026}\\
Fi & \cellcolor{diff-w-each-lang_4_0}{0.018} & \cellcolor{diff-w-each-lang_4_1}{0.066} & \cellcolor{diff-w-each-lang_4_2}{0.103} & \cellcolor{diff-w-each-lang_4_3}{0.040} & \cellcolor{diff-w-each-lang_4_4}{0.024} & \cellcolor{diff-w-each-lang_4_5}{0.071} & \cellcolor{diff-w-each-lang_4_6}{0.052} & \cellcolor{diff-w-each-lang_4_7}{0.057} & \cellcolor{diff-w-each-lang_4_8}{-0.030} & \cellcolor{diff-w-each-lang_4_9}{-0.156} & \cellcolor{diff-w-each-lang_4_10}{0.060}\\
In & \cellcolor{diff-w-each-lang_5_0}{0.066} & \cellcolor{diff-w-each-lang_5_1}{0.048} & \cellcolor{diff-w-each-lang_5_2}{0.057} & \cellcolor{diff-w-each-lang_5_3}{0.026} & \cellcolor{diff-w-each-lang_5_4}{0.036} & \cellcolor{diff-w-each-lang_5_5}{0.059} & \cellcolor{diff-w-each-lang_5_6}{0.043} & \cellcolor{diff-w-each-lang_5_7}{0.071} & \cellcolor{diff-w-each-lang_5_8}{-0.012} & \cellcolor{diff-w-each-lang_5_9}{0.130} & \cellcolor{diff-w-each-lang_5_10}{0.128}\\
Ja & \cellcolor{diff-w-each-lang_6_0}{0.051} & \cellcolor{diff-w-each-lang_6_1}{-0.074} & \cellcolor{diff-w-each-lang_6_2}{0.046} & \cellcolor{diff-w-each-lang_6_3}{-0.024} & \cellcolor{diff-w-each-lang_6_4}{0.010} & \cellcolor{diff-w-each-lang_6_5}{0.012} & \cellcolor{diff-w-each-lang_6_6}{0.015} & \cellcolor{diff-w-each-lang_6_7}{0.052} & \cellcolor{diff-w-each-lang_6_8}{-0.057} & \cellcolor{diff-w-each-lang_6_9}{-0.186} & \cellcolor{diff-w-each-lang_6_10}{0.030}\\
Ko & \cellcolor{diff-w-each-lang_7_0}{0.094} & \cellcolor{diff-w-each-lang_7_1}{-0.014} & \cellcolor{diff-w-each-lang_7_2}{0.115} & \cellcolor{diff-w-each-lang_7_3}{0.006} & \cellcolor{diff-w-each-lang_7_4}{0.051} & \cellcolor{diff-w-each-lang_7_5}{0.117} & \cellcolor{diff-w-each-lang_7_6}{0.070} & \cellcolor{diff-w-each-lang_7_7}{0.169} & \cellcolor{diff-w-each-lang_7_8}{-0.008} & \cellcolor{diff-w-each-lang_7_9}{0.104} & \cellcolor{diff-w-each-lang_7_10}{0.155}\\
Ru & \cellcolor{diff-w-each-lang_8_0}{0.094} & \cellcolor{diff-w-each-lang_8_1}{-0.003} & \cellcolor{diff-w-each-lang_8_2}{0.111} & \cellcolor{diff-w-each-lang_8_3}{0.028} & \cellcolor{diff-w-each-lang_8_4}{0.089} & \cellcolor{diff-w-each-lang_8_5}{0.064} & \cellcolor{diff-w-each-lang_8_6}{0.089} & \cellcolor{diff-w-each-lang_8_7}{0.080} & \cellcolor{diff-w-each-lang_8_8}{-0.003} & \cellcolor{diff-w-each-lang_8_9}{-0.020} & \cellcolor{diff-w-each-lang_8_10}{0.088}\\
Sw & \cellcolor{diff-w-each-lang_9_0}{0.076} & \cellcolor{diff-w-each-lang_9_1}{0.072} & \cellcolor{diff-w-each-lang_9_2}{0.137} & \cellcolor{diff-w-each-lang_9_3}{0.085} & \cellcolor{diff-w-each-lang_9_4}{0.073} & \cellcolor{diff-w-each-lang_9_5}{0.107} & \cellcolor{diff-w-each-lang_9_6}{0.048} & \cellcolor{diff-w-each-lang_9_7}{0.112} & \cellcolor{diff-w-each-lang_9_8}{0.055} & \cellcolor{diff-w-each-lang_9_9}{0.218} & \cellcolor{diff-w-each-lang_9_10}{0.103}\\
Te & \cellcolor{diff-w-each-lang_10_0}{0.063} & \cellcolor{diff-w-each-lang_10_1}{-0.003} & \cellcolor{diff-w-each-lang_10_2}{0.095} & \cellcolor{diff-w-each-lang_10_3}{-0.009} & \cellcolor{diff-w-each-lang_10_4}{0.029} & \cellcolor{diff-w-each-lang_10_5}{0.052} & \cellcolor{diff-w-each-lang_10_6}{0.045} & \cellcolor{diff-w-each-lang_10_7}{0.100} & \cellcolor{diff-w-each-lang_10_8}{0.011} & \cellcolor{diff-w-each-lang_10_9}{0.006} & \cellcolor{diff-w-each-lang_10_10}{0.014}\\
Th & \cellcolor{diff-w-each-lang_11_0}{0.082} & \cellcolor{diff-w-each-lang_11_1}{-0.024} & \cellcolor{diff-w-each-lang_11_2}{0.127} & \cellcolor{diff-w-each-lang_11_3}{0.021} & \cellcolor{diff-w-each-lang_11_4}{0.081} & \cellcolor{diff-w-each-lang_11_5}{0.066} & \cellcolor{diff-w-each-lang_11_6}{0.088} & \cellcolor{diff-w-each-lang_11_7}{0.064} & \cellcolor{diff-w-each-lang_11_8}{0.016} & \cellcolor{diff-w-each-lang_11_9}{-0.043} & \cellcolor{diff-w-each-lang_11_10}{0.023}\\

\bottomrule
\end{tabular}
}
\vspace{0.1cm}
\caption{Element-wise differences between MS pFT + FT vs.\ FT directly, answering the question: ``Does pFT always improve effectiveness?'' For in-language training, {\it yes}, as the diagonal entries are always positive.
For cross-lingual zero-shot transfer, {\it not always}, as can be seen by the negative cases off diagonal.
Positive values are highlighted in \textcolor{blue}{blue} and negative values in \textcolor{orange}{\textbf{orange}}, with saturation proportional to magnitude.
}
\label{table:delta-to-eachlang}
\vspace{-0.3cm}
\end{table}

\section{Have Model and Data!}
\label{section:case2}

\subsection{Using Multilingual BERT}
\label{sec:reuslts-case2}

{\bf Scenario (2a)}: $\mathcal{L} \in \textrm{mBERT}$ and we have data in $\mathcal{L}$.

\smallskip \noindent {\bf Recommendation: }
Start with the results of scenario (1)---to pre--fine-tune mDPR using MS MARCO passage (in English)---and then continue fine-tuning on relevance judgments in $\mathcal{L}$.
That is, always pre--fine-tune in English, even if the target language is not English, because we can benefit from cross-lingual transfer effects.
This advice is consistent with other researchers who have advocated multi-stage fine-tuning strategies~\cite{Garg_etal_AAAI2020,Xie_etal_WWW2020,ZhangXinyu_etal_ECIR2021}. $\square$

\smallskip
\noindent
{\bf Scenario (2b)}: $\mathcal{L} \in \textrm{mBERT}$ but we have no data in $\mathcal{L}$.
However, we have data in $\mathcal{K}$ and $\mathcal{K} \in \textrm{mBERT}$.
$\mathcal{K}$ may or may not be linguistically related to $\mathcal{L}$, and may or may not even be in the same script.

\medskip \noindent {\bf Recommendation: }
Start with the results of scenario (1), as above, and then continue fine-tuning on relevance judgments in $\mathcal{K}$.
That is, models may benefit from cross-lingual transfer, surprisingly, across different language families and even different scripts! $\square$

\medskip
\noindent In this section, we provide the supporting experimental evidence of how we arrive at the above recommendations.

\subsubsection{Have Data in Target Language}

Focusing on scenario (2a) specifically, rows~(5--8) in Table~\ref{table:main} start with MS MARCO pFT, i.e., row~(4), and then further fine-tune on different groups of \mrtydi training data. 
This is contrasted against rows~(9--12), without pFT (i.e., starting from the raw mBERT checkpoint).
Specifically, row~(5) and row~(9) fine-tune using the training data only in the target language $\mathcal{L}$. 
Thus, comparing these two rows answers the question:\ With data in language $\mathcal{L}$, do we still need pre--fine-tuning?
The answer appears to be, unequivocally, {\it yes}, as row~(5) beats row~(9) across all languages; these differences are statistically significant in 6 out of the 11 languages.
This means that, no matter the target language, models benefit from cross-lingual transfer from English in the pFT step.
For example, retrieval in Arabic still benefits from cross-lingual transfer with English data---despite the fact that the languages are so different (down to the written script).

A closely related question, having established the effectiveness of MS MARCO pFT + in-language FT:\ Can we further leverage cross-lingual transfer effects?
In \mrtydi, we have data for all languages, and so we experimented with MS MARCO pFT and further fine-tuning with {\it all} available data, in all languages.
These results are shown in row~(6), and we see that using data from all languages beats using data only from the target language, row~(5), in all cases except from Bengali; the gains are only statistically significant for Russian, though.
These results appear to suggest that we can not only benefit from cross-lingual transfer in the pre--fine-tuning stage, but also the further fine-tuning stage as well.
Thus, the recommendation:\ use all languages if available, because it doesn't appear to hurt, and in some cases can help.

Finally, for scenario 2(a), row~(7) reports the results of the variant where we only fine-tune with languages written in Latin script (there are no other language pairs that share the same script),
and row~(8) reports results that only fine-tune on the (Latin) target language itself and all other non-Latin-script languages
(for example, the fine-tuning languages are all but Fi, Id, Sw when the target language is English). 
For rows~(11--12), we apply the same fine-tuning treatment as rows (7--8), but start from the raw mBERT checkpoint rather than the checkpoint with pre--fine-tuning. 
We observe that scores in rows (11--12) are always higher than in row~(9), which indicates that additional training data in both in-script and out-script languages are generally beneficial when not pre--fine-tuning.
Comparing the two rows to row~(10), we observe that using all available data additionally helps.
However, this trend does not hold when pre--fine-tuning is added:\ the scores in rows~(7--8) are quite close to the scores in row~(5).

The summary here is that the models all appear to benefit from fine-tuning with all available data, although the effects are stronger without pre--fine-tuning.
This does make sense since with pFT, we are already obtaining maximal cross-language transfer effects from a large dataset (albeit from English only).
We see that row~(6) beats row~(5) for 9 of the 11 languages and has a higher overall average; this supports our overall recommendation of pre--fine-tuning and then further fine-tuning on all available data.

Putting everything together, we combine the results of the best overall multilingual dense retrieval model (pFT with MS MARCO, further fine-tuning with all data) with bag-of-words BM25; these results are shown in Table~\ref{table:main}, row~(13).
We can see, consistent with previous work, that dense and sparse retrieval models provide complementary signals, and that their combination yields effectiveness that is better than each model alone.

\subsubsection{No Data in Target Language}

Turning our attention to scenario (2b), where our target language is $\mathcal{L}$ but we only have data for $\mathcal{K}$.
The results of a matrix experiment where we consider all combinations of $\mathcal{L}, \mathcal{K}$ for all the languages in \mrtydi are shown in Table~\ref{table:mdpr-matrix}.
On the left, we fine-tune mDPR directly from the mBERT checkpoint, and on the right, we fine-tune based on an mDPR model that has already been pre--fine-tuned on MS MARCO passage.
In each case, each row represents the source of the fine-tuning data, and each column represents the target language.
The diagonal corresponds to the ``in-language'' condition described above.
As described in Section~\ref{sec:data-prep}, the amount of training data in all languages are kept the same (3,300 queries) except for Bn, Ko and Sw, which have fewer training queries and we used all training data available.

Let us consider the starting point of mDPR with MS MARCO passage pre--fine-tuning.
The research question is:\ We have training data in language $\mathcal{K}$, but it's not in our target language $\mathcal{L}$.
Should we fine-tune with it anyway?
That answer, surprisingly, is {\it yes}.
These results are shown in Table~\ref{table:delta-to-ms0shot}, which shows the element-wise difference between MS pFT + FT vs.\ MS pFT (zero-shot).
For example, mDPR (MS pFT) obtains 0.282 on Thai in a zero-shot setting, as seen in Table~\ref{table:main}, row~(4).
However, if we fine-tune further on Telugu data, we can achieve 0.436 on Thai (seen in Table~\ref{table:mdpr-matrix}, right, ``Te'' row, ``Th'' column).
This represents an improvement of 0.153 (seen in Table~\ref{table:delta-to-ms0shot}, ``Te'' row, ``Th'' column)---even though Telugu and Thai have very little in common; they are not even written in the same script!

Looking at Table~\ref{table:delta-to-ms0shot}, we see that all values are positive with the exception of a few in the English row.
This tells us that, starting with the mDPR model pre--fine-tuned on MS MARCO passages, additional fine-tuning on language $\mathcal{K}$ improves effectiveness, even if the target language is $\mathcal{L}$, where $\mathcal{L} \ne \mathcal{K}$.
The exception is retrieval in English, where we interpret the results to be a sign of overfitting, since English is already resource rich.

The above analysis uses pre--fine-tuning with MS MARCO passage as the starting point, and then further fine-tuned on other data.
However, does pFT always improve effectiveness?
That is, could it be the case that for target language $\mathcal{L}$, if we have data in $\mathcal{K}$, it's better to start from the raw mBERT checkpoint and skip pre--fine-tuning?
This question is answered by Table~\ref{table:delta-to-eachlang}, where we show the element-wise difference between MS pFT + FT vs.\ FT directly;
this corresponds to the element-wise difference between the two matrices in Table~\ref{table:mdpr-matrix}.
Along the diagonals, the answer is that pFT always helps, and this is consistent with the results above---the differences between row~(7) and row~(10) in Table~\ref{table:main}.

However, for scenario (2b), where $\mathcal{L} \neq \mathcal{K}$, the answer appears to be, not necessarily so.
For example, training on $\mathcal{K}=$ Telugu and testing on $\mathcal{L}=$ Japanese, starting with the MS pFT model confers an advantage of 0.052 over starting from the default mBERT checkpoint (seen in Table~\ref{table:delta-to-eachlang}, ``Te'' row, ``Ja'' column); 
with pFT, we obtain 0.449, but without pFT, we only reach 0.398 (see in Table~\ref{table:mdpr-matrix}).
On the flip side, training on $\mathcal{K}=$ Japanese and testing on $\mathcal{L}=$ Telugu, 
starting from the raw mBERT checkpoint beats starting with MS pFT by 0.186; with pFT, we obtain 0.429, but without pFT, we reach 0.615.
In other words, given target $\mathcal{L}$ and available relevance judgments in $\mathcal{K}$, it does not appear possible to determine, {\it a priori}, whether we should pre--fine-tune on MS MARCO first, or directly use the raw mBERT checkpoint as the starting point.

Phrased slightly differently and summarizing our recommendations for scenario (2b), for a target language $\mathcal{L}$, there does appear to exist a language $\mathcal{K}$ that provides positive cross-lingual transfer, although we are not able to generalize what characteristics lead to beneficial effects.
With pre--fine-tuning, further fine-tuning on language $\mathcal{K}$ appears not to hurt (in general), and thus forms the basis of our recommendation.
While it is true that in some cases one can achieve even better effectiveness {\it without} pre--fine-tuning (see Table~\ref{table:delta-to-eachlang}), we struggle to provide a consistent, explainable pattern, and thus we believe that pFT is a ``safer'' choice.

\subsection{Mono vs.\ Multilingual BERT}
\label{section:mono}

{\bf Scenario (2a$'$)}: In the previous exploration of scenario (2a), we assumed a starting point of mBERT.
Here, in scenario (2a$'$), we consider the case where a monolingual BERT model is available for $\mathcal{L}$.
What should we do in this case?

\smallskip \noindent {\bf Recommendation: }
Using a monolingual BERT backbone {\it can} yield a model that is more effective than using mBERT, but the monolingual model is {\it not} consistently better.
Thus, it seems ``safer'' to just use mBERT as the backbone. $\square$

\medskip \noindent Before presenting experimental results that support our recommendation, it makes sense to discuss the possible tradeoffs at work here.
Previous work has found that monolingual BERT models {\it can} be more effective than mBERT on a variety of natural language processing tasks \cite{ronnqvist-etal-2019-multilingual,martin-etal-2020-camembert,virtanen2019multilingual,armengol-estape-etal-2021-multilingual}, but the gains are \textit{not} consistent.
Obviously, the relative effectiveness depends on the quality of the monolingual model, the corpora used for pretraining, and a number of other factors.
Nevertheless, it is possible that there exist ``better'' monolingual BERT models for a target language, compared to mBERT.
However, this must be balanced against the benefits of cross-lingual transfer obtained ``for free'' from using mBERT by pre--fine-tuning on MS MARCO and leveraging datasets from other languages.
Obviously, with a monolingual (non-English) BERT, it is no longer possible to pre--fine-tune on MS MARCO.
So, the research question is:\ What's more important, overall monolingual ``in-language'' quality or cross-lingual transfer effects from pre--fine-tuning and datasets from other languages?

The answer appears to be that monolingual models {\it can} be better, but not consistently so, and thus our recommendation is that using mBERT appears to be ``safer''.
These results are shown in Table~\ref{table:main}, rows (a) to (d), where we were able to successfully train DPR in five different languages (including English) starting from a monolingual backbone (those described in Section~\ref{sec:config-model}).
We are quick to point out that these models are pretrained by different groups and may vary widely in quality, which is dependent on the pretraining corpus, hyperparameters, and a myriad of other issues.

Starting with an available monolingual BERT (in each of the target languages), we can fine-tune with just the in-language training data, shown in row~(a), all in-script training data, shown in row~(b), in-language and out-of-script training data, row~(c), and all available data, row~(d).
The results show, not surprisingly, that with monolingual BERT, fine-tuning with anything other than in-language data appears to be pointless, and in most cases actually hurt.
This makes perfect sense, as a monolingual BERT model, by definition, is not designed to process input in other languages.

The monolingual BERT vs.\ mBERT comparison is shown by row~(a) vs.\ row~(6), since with mBERT we can exploit pre--fine-tuning on MS MARCO and training data in other languages.
We see that effectiveness is comparable, and Finnish is the only language where using a monolingual model yields higher effectiveness.
In row~(f), more fully discussed in Section~\ref{section:case3}, we start with English BERT, apply pre--fine-tuning with MS MARCO and additional fine-tuning with English data.
For English retrieval, this yields the highest effectiveness, since we are leveraging both a high-quality monolingual model {\it and} exploiting (in language) pre--fine-tuning.
The difference between row~(f) and row~(5) can be attributed to mBERT vs.\ English BERT, and we can see that even in the best possible case, monolingual BERT is only marginally better than mBERT.
This supports our recommendation to always use mBERT, which we view as the ``safer'' option, and has the additional benefit of simplifying the operational aspects of model training (e.g., dealing with different tokenizers, different model sizes, etc.).\footnote{These results also say, for English retrieval, use English BERT over mBERT, which is not inconsistent with our recommendations since our focus is on multilinguality.} 

\section{Have Data, No Model!}
\label{section:case3}

\noindent {\bf Scenario (3):} $\mathcal{L} \notin \textrm{mBERT}$, but we have data in $\mathcal{L}$.

\smallskip \noindent {\bf Recommendation:} Training a dense retrieval model with an English backbone, even for a non-English target language, can be effective.
Similarly, training a retrieval model using a multilingual BERT, even one that does not include the target language, can also be effective.
Even if these dense models are not effective on their own, they appear to provide valuable additional relevance signals that can improve over bag-of-words BM25 results. 
$\square$

\medskip
\noindent The core of the recommendation here can be characterized as ``something is better than nothing'', and the results are quite surprising.
Before proceeding, however, we should first defend this as a reasonable scenario:\ We {\it do} think that it is plausible to have training data in some low-resource language that is not covered by mBERT.
In short, search existed before neural networks and pretrained transformers.
It is certainly possible for some organization to build (or even have already built) a bag-of-words search engine in some low-resource language outside the set used to pretrain mBERT, and then to use this search engine gather some relevance judgments---perhaps to train a simple learning-to-rank model.
After all, mBERT only covers around 100 languages, and there are thousands of languages we would potentially like to search in.
The amount of relevance judgments used in our experiments (as low as around 1000 examples) seems modest (see Table~\ref{tab:mr-tydi-stats}), and thus we argue that it is entirely plausible to have some training data in a target language not covered by mBERT.

What if we just started with {\it English} BERT and fine-tune with the data in $\mathcal{L}$, even though at face value this seems non-sensical, since the language could bear no similarity to English?
These results are shown in Table~\ref{table:main}:\ row~(e), without MS MARCO pFT, and row~(f), with MS MARCO pFT.
We see that, for example, training {\it English} BERT with {\it Indonesian} relevance judgments yields a model that is quite ``reasonable'' for retrieval in {\it Indonesian}.
We achieve an MRR@100 of 0.396, compared to the best possible condition, mDPR (MS pFT + all FT), which achieves 0.579.
Note that Indonesian and English are both written in the Latin script.
Even {\it across scripts}, we can achieve models that are much better than random, e.g., Arabic and Bengali, but in other cases, e.g., Telugu and Thai, the results are pretty close to garbage.
Interestingly, pre--fine-tuning English DPR on MS MARCO further improves the effectiveness across all languages, except Telugu.
This seems to indicate that, with a high-quality English DPR model, we can obtain at least a respectable dense retrieval model for many languages.

An alternative solution for this scenario is to use a multilingual BERT that does not contain the target language $\mathcal{L}$.
Here, we obviously cannot use mBERT, since all the languages in \mrtydi are in mBERT's pretraining corpus.
Instead, we used AfriBERTa \cite{ogueji-etal-2021-small}, an mBERT-style multilingual language model pretrained from scratch on 11 African languages.
Hence, all the languages in \mrtydi, except Swahili, are absent from its pretraining corpus.
Results are shown in Table~\ref{table:main}, row~(g).
Surprisingly, AfriBERTa obtains very ``reasonable'' results on most languages, despite never having been pretrained on them.
However, it seems like English BERT performs better on all languages except Telugu and Thai.
From these results, we can see that a multilingual BERT model can be used to bootstrap ``reasonable'' dense retrieval models for languages in this scenario. 

Our recommendation for scenario (3) is that if a pretrained language model does not exist for the target language, then simply using English BERT or another multilingual BERT can be helpful.
However, the astute reader might point out that in this scenario, dense retrieval models are actually worse than BM25, so what's the point of even using them?
We address this potential objection by combining the results in row~(f) with bag-of-words BM25 results, shown in Table~\ref{table:main} as row~(h).
Here, we see that the hybrid dense retrieval + BM25 scores are generally better than BM25 results alone.
Even in the case of Telugu and Thai, the two cases where English BERT yield terrible low quality output, the hybrid results are just a tiny bit worse (Te) or marginally better (Th).
Thus, it appears that even if the models trained based on our recommendation are not effective, they can still provide complementary relevance signals to improve bag-of-words BM25 in a hybrid fusion approach.
Thus, ``something is better than nothing''.

\section{Why?}
\ignore{
\begin{table}[h]
    \caption{XXX}
    \label{tab:langid}
\end{table}
}
\begin{table}[]
\resizebox{\columnwidth}{!}{
\begin{tabular}{cccccccccccc|r}
\toprule
\multicolumn{1}{l}{} & \textbf{Ar} & \textbf{Bn} & \textbf{En} & \textbf{Fi} & \textbf{Id} & \textbf{Ja} & \textbf{Ko} & \textbf{Ru} & \textbf{Sw} & \textbf{Te} & \textbf{Th} & \textbf{Total}\\
\midrule
\textbf{Ar} & \cellcolor{langid_0_0}{94.1\%} & \cellcolor{langid_1_0}{0.0\%} & \cellcolor{langid_2_0}{2.5\%} & \cellcolor{langid_3_0}{0.7\%} & \cellcolor{langid_4_0}{1.0\%} & \cellcolor{langid_5_0}{0.1\%} & \cellcolor{langid_6_0}{0.5\%} & \cellcolor{langid_7_0}{0.2\%} & \cellcolor{langid_8_0}{0.9\%} & \cellcolor{langid_9_0}{0.0\%} & \cellcolor{langid_10_0}{0.0\%} &  1,397,830 \\
\textbf{Bn} & \cellcolor{langid_0_1}{2.2\%} & \cellcolor{langid_1_1}{66.0\%} & \cellcolor{langid_2_1}{14.2\%} & \cellcolor{langid_3_1}{3.3\%} & \cellcolor{langid_4_1}{5.7\%} & \cellcolor{langid_5_1}{0.3\%} & \cellcolor{langid_6_1}{3.8\%} & \cellcolor{langid_7_1}{0.4\%} & \cellcolor{langid_8_1}{3.9\%} & \cellcolor{langid_9_1}{0.1\%} & \cellcolor{langid_10_1}{0.1\%} & 64,379 \\
\textbf{En} & \cellcolor{langid_0_2}{0.7\%} & \cellcolor{langid_1_2}{0.1\%} & \cellcolor{langid_2_2}{37.4\%} & \cellcolor{langid_3_2}{13.7\%} & \cellcolor{langid_4_2}{21.7\%} & \cellcolor{langid_5_2}{0.8\%} & \cellcolor{langid_6_2}{6.0\%} & \cellcolor{langid_7_2}{1.4\%} & \cellcolor{langid_8_2}{17.9\%} & \cellcolor{langid_9_2}{0.0\%} & \cellcolor{langid_10_2}{0.3\%} & 1,802,836 \\
\textbf{Fi} & \cellcolor{langid_0_3}{0.0\%} & \cellcolor{langid_1_3}{0.0\%} & \cellcolor{langid_2_3}{3.4\%} & \cellcolor{langid_3_3}{90.7\%} & \cellcolor{langid_4_3}{3.6\%} & \cellcolor{langid_5_3}{0.1\%} & \cellcolor{langid_6_3}{0.1\%} & \cellcolor{langid_7_3}{0.2\%} & \cellcolor{langid_8_3}{2.0\%} & \cellcolor{langid_9_3}{0.0\%} & \cellcolor{langid_10_3}{0.0\%} & 2,337,303 \\
\textbf{In} & \cellcolor{langid_0_4}{1.0\%} & \cellcolor{langid_1_4}{0.1\%} & \cellcolor{langid_2_4}{16.6\%} & \cellcolor{langid_3_4}{10.4\%} & \cellcolor{langid_4_4}{51.3\%} & \cellcolor{langid_5_4}{1.0\%} & \cellcolor{langid_6_4}{3.9\%} & \cellcolor{langid_7_4}{0.6\%} & \cellcolor{langid_8_4}{15.1\%} & \cellcolor{langid_9_4}{0.0\%} & \cellcolor{langid_10_4}{0.2\%} & 382,769 \\
\textbf{Ja} & -- & -- & -- & -- & -- & -- & -- & -- & -- & -- & -- & -- \\
\textbf{Ko} & \cellcolor{langid_0_6}{0.0\%} & \cellcolor{langid_1_6}{0.0\%} & \cellcolor{langid_2_6}{0.9\%} & \cellcolor{langid_3_6}{0.2\%} & \cellcolor{langid_4_6}{0.4\%} & \cellcolor{langid_5_6}{0.3\%} & \cellcolor{langid_6_6}{97.5\%} & \cellcolor{langid_7_6}{0.1\%} & \cellcolor{langid_8_6}{0.6\%} & \cellcolor{langid_9_6}{0.0\%} & \cellcolor{langid_10_6}{0.0\%} & 4,192,486 \\
\textbf{Ru} & \cellcolor{langid_0_7}{0.1\%} & \cellcolor{langid_1_7}{0.0\%} & \cellcolor{langid_2_7}{2.6\%} & \cellcolor{langid_3_7}{1.0\%} & \cellcolor{langid_4_7}{1.2\%} & \cellcolor{langid_5_7}{0.1\%} & \cellcolor{langid_6_7}{0.5\%} & \cellcolor{langid_7_7}{93.4\%} & \cellcolor{langid_8_7}{1.1\%} & \cellcolor{langid_9_7}{0.0\%} & \cellcolor{langid_10_7}{0.0\%} & 2,994,515 \\
\textbf{Sw} & \cellcolor{langid_0_8}{0.7\%} & \cellcolor{langid_1_8}{0.0\%} & \cellcolor{langid_2_8}{9.4\%} & \cellcolor{langid_3_8}{6.6\%} & \cellcolor{langid_4_8}{11.8\%} & \cellcolor{langid_5_8}{0.0\%} & \cellcolor{langid_6_8}{0.3\%} & \cellcolor{langid_7_8}{0.3\%} & \cellcolor{langid_8_8}{70.7\%} & \cellcolor{langid_9_8}{0.0\%} & \cellcolor{langid_10_8}{0.0\%} & 119,526 \\
\textbf{Te} & \cellcolor{langid_0_9}{0.7\%} & \cellcolor{langid_1_9}{0.2\%} & \cellcolor{langid_2_9}{16.2\%} & \cellcolor{langid_3_9}{3.5\%} & \cellcolor{langid_4_9}{7.7\%} & \cellcolor{langid_5_9}{0.1\%} & \cellcolor{langid_6_9}{0.3\%} & \cellcolor{langid_7_9}{0.3\%} & \cellcolor{langid_8_9}{6.8\%} & \cellcolor{langid_9_9}{64.3\%} & \cellcolor{langid_10_9}{0.0\%} & 65,950 \\
\textbf{Th} & \cellcolor{langid_0_10}{0.0\%} & \cellcolor{langid_1_10}{0.0\%} & \cellcolor{langid_2_10}{0.9\%} & \cellcolor{langid_3_10}{0.2\%} & \cellcolor{langid_4_10}{0.4\%} & \cellcolor{langid_5_10}{0.2\%} & \cellcolor{langid_6_10}{0.4\%} & \cellcolor{langid_7_10}{0.0\%} & \cellcolor{langid_8_10}{0.4\%} & \cellcolor{langid_9_10}{0.0\%} & \cellcolor{langid_10_10}{97.5\%} & 2,470,418 \\
\bottomrule
\end{tabular}
}
\vspace{0.1cm}
\caption{
    Analysis of unique whitespace-delimited tokens in \mrtydi.
    Each row represents a different language; we omit Japanese here because it is written without spaces.
    Each language column indicates the percentage of words belonging to that language.
    The total number of unique words in each language is reported in the rightmost column. 
    The values are highlighted \textit{row-wise}: the color gradient ranges from \textcolor{blue}{\textbf{blue}} (high values) to \textcolor{orange}{\textbf{orange}} (low values)
}
\label{tab:langid}
\vspace{-0.4cm}
\end{table}

In this paper, we have laid out a number of best practices for training multilingual dense retrieval models for monolingual retrieval, particularly in low-resource languages.
Our recommendations are supported by empirical experiments with \mrtydi, with data from eleven typologically diverse languages.
Some of our findings are quite surprising, and so the natural question is {\it why}?

We concede that we do not yet have a complete explanation of many of our counter-intuitive findings, particularly with respect to cross-lingual transfer effects across completely unrelated languages.
To be fair, though, no one in the NLP community has a completely satisfactory explanation either for many similarly surprising findings---this is an active area of research.
As we outlined in Section~\ref{section:background}, the literature describes many conflicting findings.
Nevertheless, we do have two working hypotheses that can help us begin to make sense of the observed behaviors:

\begin{itemize}[leftmargin=*]

\item Naturally occurring code-switching in Wikipedia articles provide cross-lingual anchors.

\item After mBERT tokenization, corpora from different languages share a large overlap in unique vocabulary items (i.e., subwords).

\end{itemize}

\noindent We describe these two phenomena in more detail below.

\smallskip \noindent 
{\bf Natural code-switching.}
In Wikipedia, named entities are additionally written in their ``native'' language when it is different from the language of the corpus.
This is most common in English articles, but also appears frequently in pairs of non-English languages.
For example, the Korean article about Saladin, the 12th century sultan of the Ayyubid dynasty, contains his Arabic name in the first sentence of the article.
This means that naturally occurring corpora in one language contain tokens from other languages, and we observe this occurring across all the \mrtydi languages.
Trivially, all languages share in the use of Arabic numerals as well.

We attempt to quantitatively characterize this phenomenon in the \mrtydi corpora in the following manner:
For each language, we began by tokenizing its corpus by whitespace and punctuation; we discarded Japanese from this analysis since it is written without spaces.
We applied an off-the-shelf Python language id package called \texttt{langdetect}\footnote{\url{https://github.com/Mimino666/langdetect}} to determine the language of each unique token.
These results are shown in Table~\ref{tab:langid}, where the rows represent the corpus in each language, and the column indicates the identified language.
Considering the first row, for example, on the Arabic corpus:\ We see that $94.1\%$ of unique tokens are indeed identified as Arabic.
However, there are non-trivial numbers of tokens that are identified as being from other languages as well.
For example, the Arabic article on Isaac Newton contains his name in English.
In total, we observe that about $2.5\%$ of the unique tokens in the Arabic corpus are actually identified as English.
While the results in Table~\ref{tab:langid} are a bit noisy because language id is imperfect and whitespace tokenization can be problematic for some cases,\footnote{For example, it can be more difficult to identify the language of words in Latin-script languages without context.} 
but based on manual spot checking, these results appear to be reasonable.
Regardless, this analysis conclusively shows that code-switching naturally occurs in Wikipedia.

Looking at the English row, we see that English Wikipedia exhibits the greatest diversity in code-switching ``out of English'' to other languages, e.g., Saladin to Arabic, Peter the Great to Russian, Tokugawa Ieyasu to Japanese.
Looking at the English column, we see that English is the most common language that non-English corpora code-switch ``into''.
For example, the Wikipedia page of Albert Einstein in non-English languages all mention his name in English.
While English-based cross-lingual anchors (i.e., ``pivots'') dominate, we see plenty of examples that do not involve English:\ for example, the Arabic article about Tokugawa Ieyasu contains his name in Japanese; a Russian article about the Chittoor district in India contains the native Telugu name.
Furthermore, these pivots are, for the most part, named entities---which are often the topic of natural language questions!

We believe this provides at least an initial explanation of why pre--fine-tuning with MS MARCO data works so well, even in a zero-shot setting, e.g., Table~\ref{table:main}, row~(4).
It seems reasonable to assume that mBERT has already ``internalized'' dictionaries of named entities across many languages as a natural byproduct of MLM pretraining,\footnote{We confirmed via spot-checking that this is indeed the case by forcing mBERT to perform cross-lingual prediction of masked tokens, and indeed the model was able to generate the correct named entity translation.} given that the code-switching behavior described above in essence provides parallel data (Wikipedia is part of the pretraining corpus, and it is likely that code switching is common as well in other genres of text in the pretraining corpus).
Thus, when we fine-tune in English, these cross-language anchors allow the model to transfer relevance matching capabilities across languages.
Since we have also demonstrated the existence of non-English cross-lingual anchors, this might begin to explain why fine-tuning on language $\mathcal{K}$ helps $\mathcal{L}$ (where both $\mathcal{K}$ and $\mathcal{L}$ are not English).
We further note that in QA applications, getting the named entity right is ``half the battle'' (if not more).
Consider a question about the birthdate of a person; if the model simply retrieved passages about the correct person, the results will already be ``reasonable'' (and may in fact contain the answer, just by sheer luck).

\smallskip \noindent 
{\bf Vocabulary and Parameter Sharing.}
We empirically observe that after mBERT tokenization, corpora from different languages share a large overlap in unique vocabulary items (i.e., subwords).
For example, consider the following overlap analysis, now performed in terms of mBERT tokenized subwords (just {\it subwords} hereafter).
For reference, the entire subword vocabulary space is 110k.

\begin{itemize}[leftmargin=*]

\item There are 60178 unique subwords observed in the Arabic corpus, 94836 unique subwords observed in the English corpus, and 59018 subwords that are observed in both corpora.
That is, the overlapping tokens represent $87.2\%$ of subwords in the Arabic corpus and $62.2\%$ in the English corpus. 

\item There are 33236 unique subwords observed in the Telugu corpus, 47591 unique subwords observed in the Thai corpus, and 28994 subwords that are observed in both corpora.
The overlapping tokens represents $98.1\%$ in Telugu and $60.9\%$ in Thai.

\end{itemize}

\noindent The numbers of shared subwords were a total surprise to us, and we are sure it is {\it not} a bug in our analysis code.
Furthermore, we can rule out data quality issues because the whitespace tokenization analysis above passes both sanity checks and spot checks.
The reality here is that the vast majority of shared subwords across languages are {\it not} from either language.
That is, the overlap of subwords found in both the English and Arabic corpora comprises mostly subwords {\it from other languages}!
This makes sense given the results in Table~\ref{tab:langid}, e.g., the English corpus code-switches into every other language (recall that our analysis is performed in terms of {\it unique} subwords.)

The upshot is that mBERT (during pretraining and fine-tuning) is almost never presented with clean, ``pure'', monolingual corpora.
Hence, we expect that mBERT will note common properties across multiple languages (using these cross-lingual anchors) and model the subwords jointly, thus enabling downstream cross-lingual transfer capabilities.
Furthermore, subword overlap may explain why English BERT works well even when fine-tuned on other languages in Section~\ref{section:case3}---the English corpus already contains a non-trivial amount of tokens from other languages!
We hypothesize that these shared subwords (which we show exist even in monolingual English BERT) provide anchors for downstream cross-lingual transfer.

Indeed, this explanation is consistent with the literature 
\citep{wang2019cross,conneau-etal-2020-emerging,wu-dredze-2019-beto,dufter-schutze-2020-identifying,pires-etal-2019-multilingual,lauscher-etal-2020-zero} 
which has generally agreed that higher subword overlap correlates with \textit{improved} cross-lingual ability. 
The same can be said about the model parameters; it is likely that mBERT finds similar structures among languages and models them universally.
This means that the model can easily transfer this knowledge to other unseen languages.
Since AfriBERTa is trained similarly to mBERT, it will likely have these same properties.
Coupled with token overlap, this is another possible contribution towards the impressive cross-lingual effectiveness on unseen languages discussed in Section~\ref{section:case3}.

\section{Conclusion}

Practitioners today encounter complex situations when trying to build dense retrieval models for a specific language $L$, depending on the availability of training data, models, and other factors.
In this paper, we provide guidance organized into three common scenarios.
Some of our recommendations are consistent with previous work (e.g., beneficial effects of pre--fine-tuning), but we reach some surprising conclusions as well (e.g., cross-lingual transfer across languages in different scripts).

Researchers today, on the other hand, are still trying to understand {\it why} multilingual BERT exhibits such impressive capabilities.
In contrast to most existing work on NLP tasks, our scientific contribution is to provide two possible explanations specifically for information retrieval: \textit{natural code-switching} and \textit{shared vocabulary and parameters}.
While we have not provided a completely satisfactory explanation, these ideas provide a foundation that future researchers can build on.

\section*{Acknowledgements}

This research was supported in part by the Canada First Research Excellence Fund and the Natural Sciences and Engineering Research Council (NSERC) of Canada.
Computational resources were provided in part by Compute Ontario and Compute Canada.

\bibliographystyle{ACM-Reference-Format}
\bibliography{mdpr}

\end{document}

%% file: results/colors.tex
\definecolor{table4left_0_0}{rgb}{0.122, 0.467, 0.706}
\definecolor{table4left_1_0}{rgb}{1.0, 0.628, 0.3}
\definecolor{table4left_2_0}{rgb}{1.0, 0.585, 0.218}
\definecolor{table4left_3_0}{rgb}{1.0, 0.995, 0.991}
\definecolor{table4left_4_0}{rgb}{1.0, 0.812, 0.646}
\definecolor{table4left_5_0}{rgb}{1.0, 0.783, 0.591}
\definecolor{table4left_6_0}{rgb}{1.0, 0.498, 0.055}
\definecolor{table4left_7_0}{rgb}{1.0, 0.763, 0.555}
\definecolor{table4left_8_0}{rgb}{1.0, 0.503, 0.064}
\definecolor{table4left_9_0}{rgb}{1.0, 0.937, 0.882}
\definecolor{table4left_10_0}{rgb}{1.0, 0.903, 0.818}
\definecolor{table4left_0_1}{rgb}{0.672, 0.801, 0.89}
\definecolor{table4left_1_1}{rgb}{0.122, 0.467, 0.706}
\definecolor{table4left_2_1}{rgb}{1.0, 0.498, 0.055}
\definecolor{table4left_3_1}{rgb}{1.0, 0.688, 0.412}
\definecolor{table4left_4_1}{rgb}{1.0, 0.905, 0.822}
\definecolor{table4left_5_1}{rgb}{0.58, 0.745, 0.859}
\definecolor{table4left_6_1}{rgb}{1.0, 0.8, 0.624}
\definecolor{table4left_7_1}{rgb}{0.926, 0.955, 0.975}
\definecolor{table4left_8_1}{rgb}{1.0, 0.53, 0.116}
\definecolor{table4left_9_1}{rgb}{0.481, 0.685, 0.826}
\definecolor{table4left_10_1}{rgb}{0.714, 0.827, 0.904}
\definecolor{table4left_0_2}{rgb}{0.72, 0.83, 0.906}
\definecolor{table4left_1_2}{rgb}{1.0, 0.813, 0.648}
\definecolor{table4left_2_2}{rgb}{0.122, 0.467, 0.706}
\definecolor{table4left_3_2}{rgb}{0.758, 0.853, 0.919}
\definecolor{table4left_4_2}{rgb}{0.701, 0.818, 0.9}
\definecolor{table4left_5_2}{rgb}{1.0, 0.916, 0.842}
\definecolor{table4left_6_2}{rgb}{1.0, 0.845, 0.709}
\definecolor{table4left_7_2}{rgb}{1.0, 0.965, 0.934}
\definecolor{table4left_8_2}{rgb}{1.0, 0.498, 0.055}
\definecolor{table4left_9_2}{rgb}{1.0, 0.932, 0.872}
\definecolor{table4left_10_2}{rgb}{1.0, 0.867, 0.75}
\definecolor{table4left_0_3}{rgb}{0.44, 0.66, 0.812}
\definecolor{table4left_1_3}{rgb}{1.0, 0.799, 0.622}
\definecolor{table4left_2_3}{rgb}{1.0, 0.856, 0.728}
\definecolor{table4left_3_3}{rgb}{0.122, 0.467, 0.706}
\definecolor{table4left_4_3}{rgb}{0.879, 0.927, 0.96}
\definecolor{table4left_5_3}{rgb}{1.0, 0.918, 0.846}
\definecolor{table4left_6_3}{rgb}{1.0, 0.805, 0.634}
\definecolor{table4left_7_3}{rgb}{0.934, 0.96, 0.978}
\definecolor{table4left_8_3}{rgb}{1.0, 0.498, 0.055}
\definecolor{table4left_9_3}{rgb}{1.0, 0.969, 0.941}
\definecolor{table4left_10_3}{rgb}{1.0, 0.906, 0.823}
\definecolor{table4left_0_4}{rgb}{0.58, 0.745, 0.859}
\definecolor{table4left_1_4}{rgb}{1.0, 0.6, 0.248}
\definecolor{table4left_2_4}{rgb}{1.0, 0.722, 0.476}
\definecolor{table4left_3_4}{rgb}{0.715, 0.827, 0.904}
\definecolor{table4left_4_4}{rgb}{0.122, 0.467, 0.706}
\definecolor{table4left_5_4}{rgb}{1.0, 0.677, 0.392}
\definecolor{table4left_6_4}{rgb}{1.0, 0.562, 0.175}
\definecolor{table4left_7_4}{rgb}{1.0, 0.709, 0.452}
\definecolor{table4left_8_4}{rgb}{1.0, 0.498, 0.055}
\definecolor{table4left_9_4}{rgb}{1.0, 0.735, 0.5}
\definecolor{table4left_10_4}{rgb}{1.0, 0.773, 0.573}
\definecolor{table4left_0_5}{rgb}{0.979, 0.987, 0.993}
\definecolor{table4left_1_5}{rgb}{1.0, 0.762, 0.552}
\definecolor{table4left_2_5}{rgb}{1.0, 0.677, 0.392}
\definecolor{table4left_3_5}{rgb}{1.0, 0.814, 0.649}
\definecolor{table4left_4_5}{rgb}{1.0, 0.819, 0.658}
\definecolor{table4left_5_5}{rgb}{0.122, 0.467, 0.706}
\definecolor{table4left_6_5}{rgb}{1.0, 0.738, 0.507}
\definecolor{table4left_7_5}{rgb}{1.0, 0.998, 0.996}
\definecolor{table4left_8_5}{rgb}{1.0, 0.498, 0.055}
\definecolor{table4left_9_5}{rgb}{0.724, 0.832, 0.907}
\definecolor{table4left_10_5}{rgb}{1.0, 0.984, 0.969}
\definecolor{table4left_0_6}{rgb}{0.122, 0.467, 0.706}
\definecolor{table4left_1_6}{rgb}{1.0, 0.774, 0.575}
\definecolor{table4left_2_6}{rgb}{1.0, 0.624, 0.291}
\definecolor{table4left_3_6}{rgb}{0.693, 0.813, 0.897}
\definecolor{table4left_4_6}{rgb}{0.802, 0.88, 0.934}
\definecolor{table4left_5_6}{rgb}{0.165, 0.493, 0.721}
\definecolor{table4left_6_6}{rgb}{0.144, 0.48, 0.713}
\definecolor{table4left_7_6}{rgb}{1.0, 0.586, 0.22}
\definecolor{table4left_8_6}{rgb}{1.0, 0.498, 0.055}
\definecolor{table4left_9_6}{rgb}{0.473, 0.68, 0.824}
\definecolor{table4left_10_6}{rgb}{1.0, 0.736, 0.504}
\definecolor{table4left_0_7}{rgb}{0.122, 0.467, 0.706}
\definecolor{table4left_1_7}{rgb}{1.0, 0.558, 0.168}
\definecolor{table4left_2_7}{rgb}{1.0, 0.843, 0.704}
\definecolor{table4left_3_7}{rgb}{0.528, 0.713, 0.842}
\definecolor{table4left_4_7}{rgb}{1.0, 0.91, 0.831}
\definecolor{table4left_5_7}{rgb}{0.777, 0.865, 0.925}
\definecolor{table4left_6_7}{rgb}{1.0, 0.595, 0.238}
\definecolor{table4left_7_7}{rgb}{0.541, 0.721, 0.846}
\definecolor{table4left_8_7}{rgb}{1.0, 0.498, 0.055}
\definecolor{table4left_9_7}{rgb}{1.0, 0.97, 0.944}
\definecolor{table4left_10_7}{rgb}{0.58, 0.745, 0.86}
\definecolor{table4left_0_8}{rgb}{1.0, 0.844, 0.706}
\definecolor{table4left_1_8}{rgb}{1.0, 0.696, 0.427}
\definecolor{table4left_2_8}{rgb}{1.0, 0.498, 0.055}
\definecolor{table4left_3_8}{rgb}{1.0, 0.899, 0.809}
\definecolor{table4left_4_8}{rgb}{1.0, 0.745, 0.52}
\definecolor{table4left_5_8}{rgb}{1.0, 0.739, 0.509}
\definecolor{table4left_6_8}{rgb}{1.0, 0.63, 0.303}
\definecolor{table4left_7_8}{rgb}{1.0, 0.646, 0.334}
\definecolor{table4left_8_8}{rgb}{0.122, 0.467, 0.706}
\definecolor{table4left_9_8}{rgb}{1.0, 0.63, 0.303}
\definecolor{table4left_10_8}{rgb}{1.0, 0.575, 0.2}
\definecolor{table4left_0_9}{rgb}{1.0, 0.893, 0.798}
\definecolor{table4left_1_9}{rgb}{1.0, 0.937, 0.882}
\definecolor{table4left_2_9}{rgb}{1.0, 0.52, 0.097}
\definecolor{table4left_3_9}{rgb}{0.848, 0.908, 0.949}
\definecolor{table4left_4_9}{rgb}{1.0, 0.652, 0.346}
\definecolor{table4left_5_9}{rgb}{0.86, 0.915, 0.953}
\definecolor{table4left_6_9}{rgb}{1.0, 0.539, 0.132}
\definecolor{table4left_7_9}{rgb}{1.0, 0.936, 0.879}
\definecolor{table4left_8_9}{rgb}{1.0, 0.498, 0.055}
\definecolor{table4left_9_9}{rgb}{0.122, 0.467, 0.706}
\definecolor{table4left_10_9}{rgb}{0.761, 0.855, 0.92}
\definecolor{table4left_0_10}{rgb}{0.989, 0.993, 0.996}
\definecolor{table4left_1_10}{rgb}{1.0, 0.893, 0.798}
\definecolor{table4left_2_10}{rgb}{1.0, 0.589, 0.227}
\definecolor{table4left_3_10}{rgb}{1.0, 0.779, 0.584}
\definecolor{table4left_4_10}{rgb}{1.0, 0.597, 0.241}
\definecolor{table4left_5_10}{rgb}{1.0, 0.903, 0.817}
\definecolor{table4left_6_10}{rgb}{1.0, 0.498, 0.055}
\definecolor{table4left_7_10}{rgb}{1.0, 0.651, 0.343}
\definecolor{table4left_8_10}{rgb}{1.0, 0.614, 0.273}
\definecolor{table4left_9_10}{rgb}{0.838, 0.902, 0.946}
\definecolor{table4left_10_10}{rgb}{0.122, 0.467, 0.706}

\definecolor{table4right_0_0}{rgb}{0.122, 0.467, 0.706}
\definecolor{table4right_1_0}{rgb}{1.0, 0.912, 0.834}
\definecolor{table4right_2_0}{rgb}{1.0, 0.498, 0.055}
\definecolor{table4right_3_0}{rgb}{1.0, 0.876, 0.766}
\definecolor{table4right_4_0}{rgb}{1.0, 0.922, 0.854}
\definecolor{table4right_5_0}{rgb}{1.0, 0.819, 0.659}
\definecolor{table4right_6_0}{rgb}{1.0, 0.736, 0.503}
\definecolor{table4right_7_0}{rgb}{0.964, 0.978, 0.988}
\definecolor{table4right_8_0}{rgb}{1.0, 0.648, 0.337}
\definecolor{table4right_9_0}{rgb}{0.918, 0.951, 0.973}
\definecolor{table4right_10_0}{rgb}{0.81, 0.885, 0.936}
\definecolor{table4right_0_1}{rgb}{0.401, 0.636, 0.799}
\definecolor{table4right_1_1}{rgb}{0.122, 0.467, 0.706}
\definecolor{table4right_2_1}{rgb}{1.0, 0.498, 0.055}
\definecolor{table4right_3_1}{rgb}{1.0, 0.882, 0.777}
\definecolor{table4right_4_1}{rgb}{0.928, 0.957, 0.976}
\definecolor{table4right_5_1}{rgb}{1.0, 0.877, 0.769}
\definecolor{table4right_6_1}{rgb}{1.0, 0.662, 0.363}
\definecolor{table4right_7_1}{rgb}{1.0, 0.968, 0.939}
\definecolor{table4right_8_1}{rgb}{1.0, 0.748, 0.525}
\definecolor{table4right_9_1}{rgb}{0.589, 0.751, 0.862}
\definecolor{table4right_10_1}{rgb}{0.989, 0.993, 0.996}
\definecolor{table4right_0_2}{rgb}{0.619, 0.769, 0.872}
\definecolor{table4right_1_2}{rgb}{0.917, 0.95, 0.972}
\definecolor{table4right_2_2}{rgb}{0.122, 0.467, 0.706}
\definecolor{table4right_3_2}{rgb}{0.254, 0.547, 0.75}
\definecolor{table4right_4_2}{rgb}{0.917, 0.95, 0.972}
\definecolor{table4right_5_2}{rgb}{1.0, 0.498, 0.055}
\definecolor{table4right_6_2}{rgb}{0.95, 0.97, 0.983}
\definecolor{table4right_7_2}{rgb}{0.652, 0.789, 0.883}
\definecolor{table4right_8_2}{rgb}{1.0, 0.631, 0.305}
\definecolor{table4right_9_2}{rgb}{1.0, 0.991, 0.982}
\definecolor{table4right_10_2}{rgb}{0.685, 0.809, 0.895}
\definecolor{table4right_0_3}{rgb}{0.682, 0.807, 0.894}
\definecolor{table4right_1_3}{rgb}{0.856, 0.913, 0.952}
\definecolor{table4right_2_3}{rgb}{1.0, 0.498, 0.055}
\definecolor{table4right_3_3}{rgb}{0.122, 0.467, 0.706}
\definecolor{table4right_4_3}{rgb}{0.75, 0.848, 0.916}
\definecolor{table4right_5_3}{rgb}{1.0, 0.823, 0.666}
\definecolor{table4right_6_3}{rgb}{1.0, 0.875, 0.764}
\definecolor{table4right_7_3}{rgb}{0.78, 0.867, 0.926}
\definecolor{table4right_8_3}{rgb}{0.992, 0.995, 0.997}
\definecolor{table4right_9_3}{rgb}{1.0, 0.922, 0.853}
\definecolor{table4right_10_3}{rgb}{0.985, 0.991, 0.995}
\definecolor{table4right_0_4}{rgb}{0.506, 0.7, 0.835}
\definecolor{table4right_1_4}{rgb}{0.797, 0.877, 0.932}
\definecolor{table4right_2_4}{rgb}{1.0, 0.498, 0.055}
\definecolor{table4right_3_4}{rgb}{0.601, 0.758, 0.866}
\definecolor{table4right_4_4}{rgb}{0.122, 0.467, 0.706}
\definecolor{table4right_5_4}{rgb}{1.0, 0.855, 0.727}
\definecolor{table4right_6_4}{rgb}{1.0, 0.95, 0.906}
\definecolor{table4right_7_4}{rgb}{0.652, 0.788, 0.883}
\definecolor{table4right_8_4}{rgb}{1.0, 1.0, 1.0}
\definecolor{table4right_9_4}{rgb}{1.0, 0.971, 0.945}
\definecolor{table4right_10_4}{rgb}{0.637, 0.78, 0.878}
\definecolor{table4right_0_5}{rgb}{0.881, 0.928, 0.96}
\definecolor{table4right_1_5}{rgb}{0.881, 0.928, 0.96}
\definecolor{table4right_2_5}{rgb}{1.0, 0.524, 0.103}
\definecolor{table4right_3_5}{rgb}{1.0, 0.762, 0.551}
\definecolor{table4right_4_5}{rgb}{1.0, 0.668, 0.375}
\definecolor{table4right_5_5}{rgb}{0.122, 0.467, 0.706}
\definecolor{table4right_6_5}{rgb}{0.97, 0.982, 0.99}
\definecolor{table4right_7_5}{rgb}{0.955, 0.973, 0.985}
\definecolor{table4right_8_5}{rgb}{1.0, 0.498, 0.055}
\definecolor{table4right_9_5}{rgb}{0.628, 0.774, 0.875}
\definecolor{table4right_10_5}{rgb}{0.955, 0.973, 0.985}
\definecolor{table4right_0_6}{rgb}{1.0, 0.94, 0.887}
\definecolor{table4right_1_6}{rgb}{0.96, 0.976, 0.987}
\definecolor{table4right_2_6}{rgb}{1.0, 0.498, 0.055}
\definecolor{table4right_3_6}{rgb}{0.831, 0.897, 0.943}
\definecolor{table4right_4_6}{rgb}{1.0, 0.977, 0.957}
\definecolor{table4right_5_6}{rgb}{1.0, 0.977, 0.957}
\definecolor{table4right_6_6}{rgb}{0.122, 0.467, 0.706}
\definecolor{table4right_7_6}{rgb}{0.976, 0.985, 0.992}
\definecolor{table4right_8_6}{rgb}{1.0, 0.581, 0.211}
\definecolor{table4right_9_6}{rgb}{0.766, 0.858, 0.922}
\definecolor{table4right_10_6}{rgb}{0.799, 0.878, 0.933}
\definecolor{table4right_0_7}{rgb}{0.122, 0.467, 0.706}
\definecolor{table4right_1_7}{rgb}{1.0, 0.943, 0.893}
\definecolor{table4right_2_7}{rgb}{1.0, 0.498, 0.055}
\definecolor{table4right_3_7}{rgb}{0.544, 0.723, 0.847}
\definecolor{table4right_4_7}{rgb}{0.993, 0.996, 0.998}
\definecolor{table4right_5_7}{rgb}{0.861, 0.916, 0.954}
\definecolor{table4right_6_7}{rgb}{0.254, 0.547, 0.75}
\definecolor{table4right_7_7}{rgb}{0.254, 0.547, 0.75}
\definecolor{table4right_8_7}{rgb}{1.0, 0.891, 0.794}
\definecolor{table4right_9_7}{rgb}{0.505, 0.699, 0.834}
\definecolor{table4right_10_7}{rgb}{0.505, 0.699, 0.834}
\definecolor{table4right_0_8}{rgb}{1.0, 0.869, 0.753}
\definecolor{table4right_1_8}{rgb}{1.0, 0.908, 0.827}
\definecolor{table4right_2_8}{rgb}{1.0, 0.498, 0.055}
\definecolor{table4right_3_8}{rgb}{1.0, 0.917, 0.844}
\definecolor{table4right_4_8}{rgb}{1.0, 0.884, 0.781}
\definecolor{table4right_5_8}{rgb}{1.0, 0.748, 0.526}
\definecolor{table4right_6_8}{rgb}{1.0, 0.833, 0.685}
\definecolor{table4right_7_8}{rgb}{1.0, 0.857, 0.73}
\definecolor{table4right_8_8}{rgb}{0.122, 0.467, 0.706}
\definecolor{table4right_9_8}{rgb}{1.0, 0.89, 0.793}
\definecolor{table4right_10_8}{rgb}{1.0, 0.878, 0.77}
\definecolor{table4right_0_9}{rgb}{0.886, 0.931, 0.962}
\definecolor{table4right_1_9}{rgb}{0.792, 0.874, 0.93}
\definecolor{table4right_2_9}{rgb}{1.0, 0.498, 0.055}
\definecolor{table4right_3_9}{rgb}{1.0, 0.807, 0.637}
\definecolor{table4right_4_9}{rgb}{1.0, 0.864, 0.744}
\definecolor{table4right_5_9}{rgb}{1.0, 0.749, 0.527}
\definecolor{table4right_6_9}{rgb}{1.0, 0.706, 0.447}
\definecolor{table4right_7_9}{rgb}{1.0, 0.888, 0.79}
\definecolor{table4right_8_9}{rgb}{1.0, 0.861, 0.738}
\definecolor{table4right_9_9}{rgb}{0.122, 0.467, 0.706}
\definecolor{table4right_10_9}{rgb}{0.91, 0.945, 0.97}
\definecolor{table4right_0_10}{rgb}{1.0, 0.868, 0.752}
\definecolor{table4right_1_10}{rgb}{1.0, 0.97, 0.943}
\definecolor{table4right_2_10}{rgb}{1.0, 0.498, 0.055}
\definecolor{table4right_3_10}{rgb}{1.0, 0.8, 0.623}
\definecolor{table4right_4_10}{rgb}{1.0, 0.783, 0.592}
\definecolor{table4right_5_10}{rgb}{1.0, 0.855, 0.726}
\definecolor{table4right_6_10}{rgb}{1.0, 0.748, 0.525}
\definecolor{table4right_7_10}{rgb}{1.0, 0.734, 0.499}
\definecolor{table4right_8_10}{rgb}{1.0, 0.734, 0.499}
\definecolor{table4right_9_10}{rgb}{0.962, 0.977, 0.987}
\definecolor{table4right_10_10}{rgb}{0.122, 0.467, 0.706}

\definecolor{diff-w-ms_0_0}{rgb}{0.617, 0.767, 0.872}
\definecolor{diff-w-ms_1_0}{rgb}{0.617, 0.767, 0.872}
\definecolor{diff-w-ms_2_0}{rgb}{0.816, 0.888, 0.938}
\definecolor{diff-w-ms_3_0}{rgb}{0.958, 0.974, 0.986}
\definecolor{diff-w-ms_4_0}{rgb}{0.828, 0.895, 0.942}
\definecolor{diff-w-ms_5_0}{rgb}{0.812, 0.886, 0.937}
\definecolor{diff-w-ms_6_0}{rgb}{0.847, 0.907, 0.949}
\definecolor{diff-w-ms_7_0}{rgb}{0.875, 0.924, 0.958}
\definecolor{diff-w-ms_8_0}{rgb}{0.78, 0.867, 0.926}
\definecolor{diff-w-ms_9_0}{rgb}{0.905, 0.942, 0.968}
\definecolor{diff-w-ms_10_0}{rgb}{0.77, 0.86, 0.923}
\definecolor{diff-w-ms_11_0}{rgb}{0.749, 0.847, 0.916}
\definecolor{diff-w-ms_0_1}{rgb}{0.541, 0.722, 0.846}
\definecolor{diff-w-ms_1_1}{rgb}{0.606, 0.761, 0.868}
\definecolor{diff-w-ms_2_1}{rgb}{0.541, 0.722, 0.846}
\definecolor{diff-w-ms_3_1}{rgb}{0.949, 0.969, 0.983}
\definecolor{diff-w-ms_4_1}{rgb}{0.793, 0.874, 0.931}
\definecolor{diff-w-ms_5_1}{rgb}{0.728, 0.835, 0.909}
\definecolor{diff-w-ms_6_1}{rgb}{0.796, 0.876, 0.932}
\definecolor{diff-w-ms_7_1}{rgb}{0.882, 0.929, 0.961}
\definecolor{diff-w-ms_8_1}{rgb}{0.758, 0.853, 0.919}
\definecolor{diff-w-ms_9_1}{rgb}{0.849, 0.908, 0.949}
\definecolor{diff-w-ms_10_1}{rgb}{0.649, 0.787, 0.882}
\definecolor{diff-w-ms_11_1}{rgb}{0.742, 0.843, 0.914}
\definecolor{diff-w-ms_0_2}{rgb}{0.712, 0.825, 0.904}
\definecolor{diff-w-ms_1_2}{rgb}{0.763, 0.856, 0.921}
\definecolor{diff-w-ms_2_2}{rgb}{0.794, 0.875, 0.931}
\definecolor{diff-w-ms_3_2}{rgb}{0.712, 0.825, 0.904}
\definecolor{diff-w-ms_4_2}{rgb}{0.726, 0.834, 0.908}
\definecolor{diff-w-ms_5_2}{rgb}{0.796, 0.876, 0.932}
\definecolor{diff-w-ms_6_2}{rgb}{0.896, 0.937, 0.965}
\definecolor{diff-w-ms_7_2}{rgb}{0.798, 0.877, 0.932}
\definecolor{diff-w-ms_8_2}{rgb}{0.766, 0.858, 0.922}
\definecolor{diff-w-ms_9_2}{rgb}{0.872, 0.922, 0.957}
\definecolor{diff-w-ms_10_2}{rgb}{0.807, 0.883, 0.935}
\definecolor{diff-w-ms_11_2}{rgb}{0.77, 0.86, 0.923}
\definecolor{diff-w-ms_0_3}{rgb}{0.617, 0.767, 0.872}
\definecolor{diff-w-ms_1_3}{rgb}{0.747, 0.846, 0.915}
\definecolor{diff-w-ms_2_3}{rgb}{0.787, 0.871, 0.929}
\definecolor{diff-w-ms_3_3}{rgb}{1.0, 0.986, 0.974}
\definecolor{diff-w-ms_4_3}{rgb}{0.617, 0.767, 0.872}
\definecolor{diff-w-ms_5_3}{rgb}{0.763, 0.856, 0.921}
\definecolor{diff-w-ms_6_3}{rgb}{0.893, 0.935, 0.964}
\definecolor{diff-w-ms_7_3}{rgb}{0.872, 0.922, 0.957}
\definecolor{diff-w-ms_8_3}{rgb}{0.77, 0.86, 0.923}
\definecolor{diff-w-ms_9_3}{rgb}{0.819, 0.89, 0.939}
\definecolor{diff-w-ms_10_3}{rgb}{0.852, 0.91, 0.951}
\definecolor{diff-w-ms_11_3}{rgb}{0.817, 0.889, 0.939}
\definecolor{diff-w-ms_0_4}{rgb}{0.678, 0.805, 0.892}
\definecolor{diff-w-ms_1_4}{rgb}{0.772, 0.861, 0.924}
\definecolor{diff-w-ms_2_4}{rgb}{0.842, 0.904, 0.947}
\definecolor{diff-w-ms_3_4}{rgb}{1.0, 0.941, 0.888}
\definecolor{diff-w-ms_4_4}{rgb}{0.794, 0.875, 0.931}
\definecolor{diff-w-ms_5_4}{rgb}{0.678, 0.805, 0.892}
\definecolor{diff-w-ms_6_4}{rgb}{0.953, 0.971, 0.984}
\definecolor{diff-w-ms_7_4}{rgb}{0.912, 0.947, 0.971}
\definecolor{diff-w-ms_8_4}{rgb}{0.807, 0.883, 0.935}
\definecolor{diff-w-ms_9_4}{rgb}{0.891, 0.934, 0.964}
\definecolor{diff-w-ms_10_4}{rgb}{0.903, 0.941, 0.968}
\definecolor{diff-w-ms_11_4}{rgb}{0.803, 0.881, 0.934}
\definecolor{diff-w-ms_0_5}{rgb}{0.703, 0.82, 0.901}
\definecolor{diff-w-ms_1_5}{rgb}{0.791, 0.873, 0.93}
\definecolor{diff-w-ms_2_5}{rgb}{0.793, 0.874, 0.931}
\definecolor{diff-w-ms_3_5}{rgb}{0.903, 0.941, 0.968}
\definecolor{diff-w-ms_4_5}{rgb}{0.856, 0.913, 0.952}
\definecolor{diff-w-ms_5_5}{rgb}{0.875, 0.924, 0.958}
\definecolor{diff-w-ms_6_5}{rgb}{0.703, 0.82, 0.901}
\definecolor{diff-w-ms_7_5}{rgb}{0.803, 0.881, 0.934}
\definecolor{diff-w-ms_8_5}{rgb}{0.8, 0.878, 0.933}
\definecolor{diff-w-ms_9_5}{rgb}{0.91, 0.946, 0.97}
\definecolor{diff-w-ms_10_5}{rgb}{0.761, 0.855, 0.92}
\definecolor{diff-w-ms_11_5}{rgb}{0.801, 0.879, 0.934}
\definecolor{diff-w-ms_0_6}{rgb}{0.726, 0.834, 0.908}
\definecolor{diff-w-ms_1_6}{rgb}{0.833, 0.899, 0.944}
\definecolor{diff-w-ms_2_6}{rgb}{0.817, 0.889, 0.939}
\definecolor{diff-w-ms_3_6}{rgb}{0.917, 0.95, 0.972}
\definecolor{diff-w-ms_4_6}{rgb}{0.803, 0.881, 0.934}
\definecolor{diff-w-ms_5_6}{rgb}{0.826, 0.894, 0.942}
\definecolor{diff-w-ms_6_6}{rgb}{0.826, 0.894, 0.942}
\definecolor{diff-w-ms_7_6}{rgb}{0.726, 0.834, 0.908}
\definecolor{diff-w-ms_8_6}{rgb}{0.819, 0.89, 0.939}
\definecolor{diff-w-ms_9_6}{rgb}{0.902, 0.94, 0.967}
\definecolor{diff-w-ms_10_6}{rgb}{0.796, 0.876, 0.932}
\definecolor{diff-w-ms_11_6}{rgb}{0.8, 0.878, 0.933}
\definecolor{diff-w-ms_0_7}{rgb}{0.715, 0.827, 0.905}
\definecolor{diff-w-ms_1_7}{rgb}{0.698, 0.817, 0.899}
\definecolor{diff-w-ms_2_7}{rgb}{0.828, 0.895, 0.942}
\definecolor{diff-w-ms_3_7}{rgb}{0.931, 0.958, 0.977}
\definecolor{diff-w-ms_4_7}{rgb}{0.754, 0.851, 0.918}
\definecolor{diff-w-ms_5_7}{rgb}{0.814, 0.887, 0.938}
\definecolor{diff-w-ms_6_7}{rgb}{0.796, 0.876, 0.932}
\definecolor{diff-w-ms_7_7}{rgb}{0.715, 0.827, 0.905}
\definecolor{diff-w-ms_8_7}{rgb}{0.715, 0.827, 0.905}
\definecolor{diff-w-ms_9_7}{rgb}{0.838, 0.902, 0.946}
\definecolor{diff-w-ms_10_7}{rgb}{0.749, 0.847, 0.916}
\definecolor{diff-w-ms_11_7}{rgb}{0.749, 0.847, 0.916}
\definecolor{diff-w-ms_0_8}{rgb}{0.533, 0.716, 0.844}
\definecolor{diff-w-ms_1_8}{rgb}{0.902, 0.94, 0.967}
\definecolor{diff-w-ms_2_8}{rgb}{0.879, 0.926, 0.959}
\definecolor{diff-w-ms_3_8}{rgb}{1.0, 0.933, 0.873}
\definecolor{diff-w-ms_4_8}{rgb}{0.874, 0.923, 0.958}
\definecolor{diff-w-ms_5_8}{rgb}{0.893, 0.935, 0.964}
\definecolor{diff-w-ms_6_8}{rgb}{0.972, 0.983, 0.991}
\definecolor{diff-w-ms_7_8}{rgb}{0.923, 0.953, 0.974}
\definecolor{diff-w-ms_8_8}{rgb}{0.909, 0.945, 0.969}
\definecolor{diff-w-ms_9_8}{rgb}{0.533, 0.716, 0.844}
\definecolor{diff-w-ms_10_8}{rgb}{0.889, 0.933, 0.963}
\definecolor{diff-w-ms_11_8}{rgb}{0.896, 0.937, 0.965}
\definecolor{diff-w-ms_0_9}{rgb}{0.122, 0.467, 0.706}
\definecolor{diff-w-ms_1_9}{rgb}{0.559, 0.732, 0.852}
\definecolor{diff-w-ms_2_9}{rgb}{0.506, 0.7, 0.835}
\definecolor{diff-w-ms_3_9}{rgb}{1.0, 0.92, 0.849}
\definecolor{diff-w-ms_4_9}{rgb}{0.824, 0.893, 0.941}
\definecolor{diff-w-ms_5_9}{rgb}{0.766, 0.858, 0.922}
\definecolor{diff-w-ms_6_9}{rgb}{0.884, 0.93, 0.961}
\definecolor{diff-w-ms_7_9}{rgb}{0.928, 0.956, 0.976}
\definecolor{diff-w-ms_8_9}{rgb}{0.742, 0.843, 0.914}
\definecolor{diff-w-ms_9_9}{rgb}{0.77, 0.86, 0.923}
\definecolor{diff-w-ms_10_9}{rgb}{0.122, 0.467, 0.706}
\definecolor{diff-w-ms_11_9}{rgb}{0.575, 0.742, 0.858}
\definecolor{diff-w-ms_0_10}{rgb}{0.422, 0.649, 0.806}
\definecolor{diff-w-ms_1_10}{rgb}{0.83, 0.897, 0.943}
\definecolor{diff-w-ms_2_10}{rgb}{0.765, 0.857, 0.921}
\definecolor{diff-w-ms_3_10}{rgb}{1.0, 0.962, 0.928}
\definecolor{diff-w-ms_4_10}{rgb}{0.872, 0.922, 0.957}
\definecolor{diff-w-ms_5_10}{rgb}{0.882, 0.929, 0.961}
\definecolor{diff-w-ms_6_10}{rgb}{0.838, 0.902, 0.946}
\definecolor{diff-w-ms_7_10}{rgb}{0.905, 0.942, 0.968}
\definecolor{diff-w-ms_8_10}{rgb}{0.916, 0.949, 0.972}
\definecolor{diff-w-ms_9_10}{rgb}{0.916, 0.949, 0.972}
\definecolor{diff-w-ms_10_10}{rgb}{0.731, 0.837, 0.91}
\definecolor{diff-w-ms_11_10}{rgb}{0.422, 0.649, 0.806}

\definecolor{diff-w-each-lang_0_0}{rgb}{0.942, 0.965, 0.981}
\definecolor{diff-w-each-lang_1_0}{rgb}{0.942, 0.965, 0.981}
\definecolor{diff-w-each-lang_2_0}{rgb}{0.823, 0.892, 0.941}
\definecolor{diff-w-each-lang_3_0}{rgb}{0.949, 0.969, 0.983}
\definecolor{diff-w-each-lang_4_0}{rgb}{0.968, 0.981, 0.989}
\definecolor{diff-w-each-lang_5_0}{rgb}{0.884, 0.93, 0.961}
\definecolor{diff-w-each-lang_6_0}{rgb}{0.91, 0.946, 0.97}
\definecolor{diff-w-each-lang_7_0}{rgb}{0.835, 0.9, 0.945}
\definecolor{diff-w-each-lang_8_0}{rgb}{0.835, 0.9, 0.945}
\definecolor{diff-w-each-lang_9_0}{rgb}{0.866, 0.919, 0.955}
\definecolor{diff-w-each-lang_10_0}{rgb}{0.889, 0.933, 0.963}
\definecolor{diff-w-each-lang_11_0}{rgb}{0.856, 0.913, 0.952}
\definecolor{diff-w-each-lang_0_1}{rgb}{0.989, 0.994, 0.996}
\definecolor{diff-w-each-lang_1_1}{rgb}{0.917, 0.95, 0.972}
\definecolor{diff-w-each-lang_2_1}{rgb}{0.989, 0.994, 0.996}
\definecolor{diff-w-each-lang_3_1}{rgb}{0.96, 0.975, 0.986}
\definecolor{diff-w-each-lang_4_1}{rgb}{0.884, 0.93, 0.961}
\definecolor{diff-w-each-lang_5_1}{rgb}{0.916, 0.949, 0.972}
\definecolor{diff-w-each-lang_6_1}{rgb}{1.0, 0.926, 0.86}
\definecolor{diff-w-each-lang_7_1}{rgb}{1.0, 0.986, 0.974}
\definecolor{diff-w-each-lang_8_1}{rgb}{1.0, 0.997, 0.994}
\definecolor{diff-w-each-lang_9_1}{rgb}{0.874, 0.923, 0.958}
\definecolor{diff-w-each-lang_10_1}{rgb}{1.0, 0.997, 0.994}
\definecolor{diff-w-each-lang_11_1}{rgb}{1.0, 0.976, 0.955}
\definecolor{diff-w-each-lang_0_2}{rgb}{0.923, 0.953, 0.974}
\definecolor{diff-w-each-lang_1_2}{rgb}{0.865, 0.918, 0.955}
\definecolor{diff-w-each-lang_2_2}{rgb}{0.784, 0.869, 0.928}
\definecolor{diff-w-each-lang_3_2}{rgb}{0.923, 0.953, 0.974}
\definecolor{diff-w-each-lang_4_2}{rgb}{0.819, 0.89, 0.939}
\definecolor{diff-w-each-lang_5_2}{rgb}{0.9, 0.939, 0.966}
\definecolor{diff-w-each-lang_6_2}{rgb}{0.919, 0.951, 0.973}
\definecolor{diff-w-each-lang_7_2}{rgb}{0.798, 0.877, 0.932}
\definecolor{diff-w-each-lang_8_2}{rgb}{0.805, 0.882, 0.935}
\definecolor{diff-w-each-lang_9_2}{rgb}{0.759, 0.854, 0.919}
\definecolor{diff-w-each-lang_10_2}{rgb}{0.833, 0.899, 0.944}
\definecolor{diff-w-each-lang_11_2}{rgb}{0.777, 0.865, 0.925}
\definecolor{diff-w-each-lang_0_3}{rgb}{0.93, 0.957, 0.976}
\definecolor{diff-w-each-lang_1_3}{rgb}{1.0, 0.995, 0.991}
\definecolor{diff-w-each-lang_2_3}{rgb}{0.903, 0.941, 0.968}
\definecolor{diff-w-each-lang_3_3}{rgb}{1.0, 0.911, 0.832}
\definecolor{diff-w-each-lang_4_3}{rgb}{0.93, 0.957, 0.976}
\definecolor{diff-w-each-lang_5_3}{rgb}{0.954, 0.972, 0.985}
\definecolor{diff-w-each-lang_6_3}{rgb}{1.0, 0.976, 0.955}
\definecolor{diff-w-each-lang_7_3}{rgb}{0.989, 0.994, 0.996}
\definecolor{diff-w-each-lang_8_3}{rgb}{0.951, 0.97, 0.984}
\definecolor{diff-w-each-lang_9_3}{rgb}{0.851, 0.909, 0.95}
\definecolor{diff-w-each-lang_10_3}{rgb}{1.0, 0.991, 0.983}
\definecolor{diff-w-each-lang_11_3}{rgb}{0.963, 0.978, 0.988}
\definecolor{diff-w-each-lang_0_4}{rgb}{0.937, 0.962, 0.979}
\definecolor{diff-w-each-lang_1_4}{rgb}{0.956, 0.973, 0.985}
\definecolor{diff-w-each-lang_2_4}{rgb}{0.851, 0.909, 0.95}
\definecolor{diff-w-each-lang_3_4}{rgb}{1.0, 0.917, 0.843}
\definecolor{diff-w-each-lang_4_4}{rgb}{0.958, 0.974, 0.986}
\definecolor{diff-w-each-lang_5_4}{rgb}{0.937, 0.962, 0.979}
\definecolor{diff-w-each-lang_6_4}{rgb}{0.982, 0.989, 0.994}
\definecolor{diff-w-each-lang_7_4}{rgb}{0.91, 0.946, 0.97}
\definecolor{diff-w-each-lang_8_4}{rgb}{0.844, 0.905, 0.948}
\definecolor{diff-w-each-lang_9_4}{rgb}{0.872, 0.922, 0.957}
\definecolor{diff-w-each-lang_10_4}{rgb}{0.949, 0.969, 0.983}
\definecolor{diff-w-each-lang_11_4}{rgb}{0.858, 0.914, 0.952}
\definecolor{diff-w-each-lang_0_5}{rgb}{0.979, 0.987, 0.993}
\definecolor{diff-w-each-lang_1_5}{rgb}{0.886, 0.931, 0.962}
\definecolor{diff-w-each-lang_2_5}{rgb}{0.793, 0.874, 0.931}
\definecolor{diff-w-each-lang_3_5}{rgb}{0.874, 0.923, 0.958}
\definecolor{diff-w-each-lang_4_5}{rgb}{0.875, 0.924, 0.958}
\definecolor{diff-w-each-lang_5_5}{rgb}{0.896, 0.937, 0.965}
\definecolor{diff-w-each-lang_6_5}{rgb}{0.979, 0.987, 0.993}
\definecolor{diff-w-each-lang_7_5}{rgb}{0.794, 0.875, 0.931}
\definecolor{diff-w-each-lang_8_5}{rgb}{0.888, 0.932, 0.962}
\definecolor{diff-w-each-lang_9_5}{rgb}{0.812, 0.886, 0.937}
\definecolor{diff-w-each-lang_10_5}{rgb}{0.909, 0.945, 0.969}
\definecolor{diff-w-each-lang_11_5}{rgb}{0.884, 0.93, 0.961}
\definecolor{diff-w-each-lang_0_6}{rgb}{0.877, 0.925, 0.959}
\definecolor{diff-w-each-lang_1_6}{rgb}{0.986, 0.991, 0.995}
\definecolor{diff-w-each-lang_2_6}{rgb}{0.868, 0.92, 0.956}
\definecolor{diff-w-each-lang_3_6}{rgb}{0.947, 0.968, 0.982}
\definecolor{diff-w-each-lang_4_6}{rgb}{0.909, 0.945, 0.969}
\definecolor{diff-w-each-lang_5_6}{rgb}{0.924, 0.954, 0.975}
\definecolor{diff-w-each-lang_6_6}{rgb}{0.974, 0.984, 0.991}
\definecolor{diff-w-each-lang_7_6}{rgb}{0.877, 0.925, 0.959}
\definecolor{diff-w-each-lang_8_6}{rgb}{0.844, 0.905, 0.948}
\definecolor{diff-w-each-lang_9_6}{rgb}{0.916, 0.949, 0.972}
\definecolor{diff-w-each-lang_10_6}{rgb}{0.921, 0.952, 0.974}
\definecolor{diff-w-each-lang_11_6}{rgb}{0.845, 0.906, 0.948}
\definecolor{diff-w-each-lang_0_7}{rgb}{0.859, 0.915, 0.953}
\definecolor{diff-w-each-lang_1_7}{rgb}{0.898, 0.938, 0.966}
\definecolor{diff-w-each-lang_2_7}{rgb}{0.807, 0.883, 0.935}
\definecolor{diff-w-each-lang_3_7}{rgb}{0.977, 0.986, 0.992}
\definecolor{diff-w-each-lang_4_7}{rgb}{0.9, 0.939, 0.966}
\definecolor{diff-w-each-lang_5_7}{rgb}{0.875, 0.924, 0.958}
\definecolor{diff-w-each-lang_6_7}{rgb}{0.909, 0.945, 0.969}
\definecolor{diff-w-each-lang_7_7}{rgb}{0.703, 0.82, 0.901}
\definecolor{diff-w-each-lang_8_7}{rgb}{0.859, 0.915, 0.953}
\definecolor{diff-w-each-lang_9_7}{rgb}{0.803, 0.881, 0.934}
\definecolor{diff-w-each-lang_10_7}{rgb}{0.824, 0.893, 0.941}
\definecolor{diff-w-each-lang_11_7}{rgb}{0.888, 0.932, 0.962}
\definecolor{diff-w-each-lang_0_8}{rgb}{0.903, 0.941, 0.968}
\definecolor{diff-w-each-lang_1_8}{rgb}{1.0, 0.965, 0.934}
\definecolor{diff-w-each-lang_2_8}{rgb}{0.989, 0.994, 0.996}
\definecolor{diff-w-each-lang_3_8}{rgb}{1.0, 0.905, 0.82}
\definecolor{diff-w-each-lang_4_8}{rgb}{1.0, 0.97, 0.943}
\definecolor{diff-w-each-lang_5_8}{rgb}{1.0, 0.988, 0.977}
\definecolor{diff-w-each-lang_6_8}{rgb}{1.0, 0.943, 0.892}
\definecolor{diff-w-each-lang_7_8}{rgb}{1.0, 0.992, 0.985}
\definecolor{diff-w-each-lang_8_8}{rgb}{1.0, 0.997, 0.994}
\definecolor{diff-w-each-lang_9_8}{rgb}{0.903, 0.941, 0.968}
\definecolor{diff-w-each-lang_10_8}{rgb}{0.981, 0.988, 0.994}
\definecolor{diff-w-each-lang_11_8}{rgb}{0.972, 0.983, 0.991}
\definecolor{diff-w-each-lang_0_9}{rgb}{0.989, 0.994, 0.996}
\definecolor{diff-w-each-lang_1_9}{rgb}{0.812, 0.886, 0.937}
\definecolor{diff-w-each-lang_2_9}{rgb}{0.803, 0.881, 0.934}
\definecolor{diff-w-each-lang_3_9}{rgb}{1.0, 0.994, 0.989}
\definecolor{diff-w-each-lang_4_9}{rgb}{1.0, 0.843, 0.705}
\definecolor{diff-w-each-lang_5_9}{rgb}{0.772, 0.861, 0.924}
\definecolor{diff-w-each-lang_6_9}{rgb}{1.0, 0.813, 0.648}
\definecolor{diff-w-each-lang_7_9}{rgb}{0.817, 0.889, 0.939}
\definecolor{diff-w-each-lang_8_9}{rgb}{1.0, 0.98, 0.962}
\definecolor{diff-w-each-lang_9_9}{rgb}{0.617, 0.767, 0.872}
\definecolor{diff-w-each-lang_10_9}{rgb}{0.989, 0.994, 0.996}
\definecolor{diff-w-each-lang_11_9}{rgb}{1.0, 0.957, 0.919}
\definecolor{diff-w-each-lang_0_10}{rgb}{0.96, 0.975, 0.986}
\definecolor{diff-w-each-lang_1_10}{rgb}{1.0, 0.993, 0.987}
\definecolor{diff-w-each-lang_2_10}{rgb}{0.866, 0.919, 0.955}
\definecolor{diff-w-each-lang_3_10}{rgb}{0.954, 0.972, 0.985}
\definecolor{diff-w-each-lang_4_10}{rgb}{0.895, 0.936, 0.965}
\definecolor{diff-w-each-lang_5_10}{rgb}{0.775, 0.863, 0.925}
\definecolor{diff-w-each-lang_6_10}{rgb}{0.947, 0.968, 0.982}
\definecolor{diff-w-each-lang_7_10}{rgb}{0.728, 0.835, 0.909}
\definecolor{diff-w-each-lang_8_10}{rgb}{0.845, 0.906, 0.948}
\definecolor{diff-w-each-lang_9_10}{rgb}{0.819, 0.89, 0.939}
\definecolor{diff-w-each-lang_10_10}{rgb}{0.975, 0.985, 0.992}
\definecolor{diff-w-each-lang_11_10}{rgb}{0.96, 0.975, 0.986}

\definecolor{langid_0_0}{rgb}{0.122, 0.467, 0.706}
\definecolor{langid_1_0}{rgb}{1.0, 0.498, 0.055}
\definecolor{langid_2_0}{rgb}{1.0, 0.525, 0.105}
\definecolor{langid_3_0}{rgb}{1.0, 0.506, 0.069}
\definecolor{langid_4_0}{rgb}{1.0, 0.509, 0.075}
\definecolor{langid_5_0}{rgb}{1.0, 0.499, 0.057}
\definecolor{langid_6_0}{rgb}{1.0, 0.503, 0.065}
\definecolor{langid_7_0}{rgb}{1.0, 0.5, 0.059}
\definecolor{langid_8_0}{rgb}{1.0, 0.508, 0.073}
\definecolor{langid_9_0}{rgb}{1.0, 0.498, 0.055}
\definecolor{langid_10_0}{rgb}{1.0, 0.498, 0.055}
\definecolor{langid_0_1}{rgb}{1.0, 0.53, 0.115}
\definecolor{langid_1_1}{rgb}{0.122, 0.467, 0.706}
\definecolor{langid_2_1}{rgb}{1.0, 0.713, 0.459}
\definecolor{langid_3_1}{rgb}{1.0, 0.547, 0.147}
\definecolor{langid_4_1}{rgb}{1.0, 0.583, 0.216}
\definecolor{langid_5_1}{rgb}{1.0, 0.501, 0.061}
\definecolor{langid_6_1}{rgb}{1.0, 0.554, 0.161}
\definecolor{langid_7_1}{rgb}{1.0, 0.503, 0.064}
\definecolor{langid_8_1}{rgb}{1.0, 0.556, 0.164}
\definecolor{langid_9_1}{rgb}{1.0, 0.498, 0.055}
\definecolor{langid_10_1}{rgb}{1.0, 0.498, 0.055}
\definecolor{langid_0_2}{rgb}{1.0, 0.517, 0.09}
\definecolor{langid_1_2}{rgb}{1.0, 0.501, 0.06}
\definecolor{langid_2_2}{rgb}{0.122, 0.467, 0.706}
\definecolor{langid_3_2}{rgb}{1.0, 0.866, 0.747}
\definecolor{langid_4_2}{rgb}{0.859, 0.914, 0.953}
\definecolor{langid_5_2}{rgb}{1.0, 0.52, 0.095}
\definecolor{langid_6_2}{rgb}{1.0, 0.659, 0.358}
\definecolor{langid_7_2}{rgb}{1.0, 0.536, 0.126}
\definecolor{langid_8_2}{rgb}{1.0, 0.979, 0.96}
\definecolor{langid_9_2}{rgb}{1.0, 0.498, 0.055}
\definecolor{langid_10_2}{rgb}{1.0, 0.506, 0.07}
\definecolor{langid_0_3}{rgb}{1.0, 0.498, 0.055}
\definecolor{langid_1_3}{rgb}{1.0, 0.498, 0.055}
\definecolor{langid_2_3}{rgb}{1.0, 0.536, 0.126}
\definecolor{langid_3_3}{rgb}{0.122, 0.467, 0.706}
\definecolor{langid_4_3}{rgb}{1.0, 0.538, 0.13}
\definecolor{langid_5_3}{rgb}{1.0, 0.499, 0.057}
\definecolor{langid_6_3}{rgb}{1.0, 0.499, 0.057}
\definecolor{langid_7_3}{rgb}{1.0, 0.5, 0.059}
\definecolor{langid_8_3}{rgb}{1.0, 0.52, 0.097}
\definecolor{langid_9_3}{rgb}{1.0, 0.498, 0.055}
\definecolor{langid_10_3}{rgb}{1.0, 0.498, 0.055}
\definecolor{langid_0_4}{rgb}{1.0, 0.518, 0.092}
\definecolor{langid_1_4}{rgb}{1.0, 0.5, 0.059}
\definecolor{langid_2_4}{rgb}{1.0, 0.823, 0.667}
\definecolor{langid_3_4}{rgb}{1.0, 0.702, 0.438}
\definecolor{langid_4_4}{rgb}{0.122, 0.467, 0.706}
\definecolor{langid_5_4}{rgb}{1.0, 0.518, 0.092}
\definecolor{langid_6_4}{rgb}{1.0, 0.574, 0.199}
\definecolor{langid_7_4}{rgb}{1.0, 0.51, 0.077}
\definecolor{langid_8_4}{rgb}{1.0, 0.794, 0.611}
\definecolor{langid_9_4}{rgb}{1.0, 0.498, 0.055}
\definecolor{langid_10_4}{rgb}{1.0, 0.502, 0.062}
\definecolor{langid_0_5}{rgb}{1.0, 0.498, 0.055}
\definecolor{langid_1_5}{rgb}{1.0, 0.498, 0.055}
\definecolor{langid_2_5}{rgb}{1.0, 0.5, 0.059}
\definecolor{langid_3_5}{rgb}{1.0, 0.499, 0.057}
\definecolor{langid_4_5}{rgb}{1.0, 0.499, 0.057}
\definecolor{langid_5_5}{rgb}{0.122, 0.467, 0.706}
\definecolor{langid_6_5}{rgb}{1.0, 0.591, 0.229}
\definecolor{langid_7_5}{rgb}{1.0, 0.498, 0.055}
\definecolor{langid_8_5}{rgb}{1.0, 0.499, 0.057}
\definecolor{langid_9_5}{rgb}{1.0, 0.498, 0.055}
\definecolor{langid_10_5}{rgb}{1.0, 0.498, 0.055}
\definecolor{langid_0_6}{rgb}{1.0, 0.498, 0.055}
\definecolor{langid_1_6}{rgb}{1.0, 0.498, 0.055}
\definecolor{langid_2_6}{rgb}{1.0, 0.507, 0.072}
\definecolor{langid_3_6}{rgb}{1.0, 0.5, 0.059}
\definecolor{langid_4_6}{rgb}{1.0, 0.502, 0.063}
\definecolor{langid_5_6}{rgb}{1.0, 0.501, 0.061}
\definecolor{langid_6_6}{rgb}{0.122, 0.467, 0.706}
\definecolor{langid_7_6}{rgb}{1.0, 0.499, 0.057}
\definecolor{langid_8_6}{rgb}{1.0, 0.504, 0.067}
\definecolor{langid_9_6}{rgb}{1.0, 0.498, 0.055}
\definecolor{langid_10_6}{rgb}{1.0, 0.498, 0.055}
\definecolor{langid_0_7}{rgb}{1.0, 0.499, 0.057}
\definecolor{langid_1_7}{rgb}{1.0, 0.498, 0.055}
\definecolor{langid_2_7}{rgb}{1.0, 0.526, 0.108}
\definecolor{langid_3_7}{rgb}{1.0, 0.509, 0.075}
\definecolor{langid_4_7}{rgb}{1.0, 0.511, 0.079}
\definecolor{langid_5_7}{rgb}{1.0, 0.499, 0.057}
\definecolor{langid_6_7}{rgb}{1.0, 0.503, 0.065}
\definecolor{langid_7_7}{rgb}{0.122, 0.467, 0.706}
\definecolor{langid_8_7}{rgb}{1.0, 0.51, 0.077}
\definecolor{langid_9_7}{rgb}{1.0, 0.498, 0.055}
\definecolor{langid_10_7}{rgb}{1.0, 0.498, 0.055}
\definecolor{langid_0_8}{rgb}{1.0, 0.508, 0.074}
\definecolor{langid_1_8}{rgb}{1.0, 0.498, 0.055}
\definecolor{langid_2_8}{rgb}{1.0, 0.632, 0.306}
\definecolor{langid_3_8}{rgb}{1.0, 0.592, 0.231}
\definecolor{langid_4_8}{rgb}{1.0, 0.666, 0.37}
\definecolor{langid_5_8}{rgb}{1.0, 0.498, 0.055}
\definecolor{langid_6_8}{rgb}{1.0, 0.502, 0.063}
\definecolor{langid_7_8}{rgb}{1.0, 0.502, 0.063}
\definecolor{langid_8_8}{rgb}{0.122, 0.467, 0.706}
\definecolor{langid_9_8}{rgb}{1.0, 0.498, 0.055}
\definecolor{langid_10_8}{rgb}{1.0, 0.498, 0.055}
\definecolor{langid_0_9}{rgb}{1.0, 0.509, 0.075}
\definecolor{langid_1_9}{rgb}{1.0, 0.501, 0.061}
\definecolor{langid_2_9}{rgb}{1.0, 0.751, 0.531}
\definecolor{langid_3_9}{rgb}{1.0, 0.553, 0.158}
\definecolor{langid_4_9}{rgb}{1.0, 0.618, 0.281}
\definecolor{langid_5_9}{rgb}{1.0, 0.5, 0.058}
\definecolor{langid_6_9}{rgb}{1.0, 0.503, 0.064}
\definecolor{langid_7_9}{rgb}{1.0, 0.503, 0.064}
\definecolor{langid_8_9}{rgb}{1.0, 0.604, 0.255}
\definecolor{langid_9_9}{rgb}{0.122, 0.467, 0.706}
\definecolor{langid_10_9}{rgb}{1.0, 0.498, 0.055}
\definecolor{langid_0_10}{rgb}{1.0, 0.498, 0.055}
\definecolor{langid_1_10}{rgb}{1.0, 0.498, 0.055}
\definecolor{langid_2_10}{rgb}{1.0, 0.507, 0.072}
\definecolor{langid_3_10}{rgb}{1.0, 0.5, 0.059}
\definecolor{langid_4_10}{rgb}{1.0, 0.502, 0.063}
\definecolor{langid_5_10}{rgb}{1.0, 0.5, 0.059}
\definecolor{langid_6_10}{rgb}{1.0, 0.502, 0.063}
\definecolor{langid_7_10}{rgb}{1.0, 0.498, 0.055}
\definecolor{langid_8_10}{rgb}{1.0, 0.502, 0.063}
\definecolor{langid_9_10}{rgb}{1.0, 0.498, 0.055}
\definecolor{langid_10_10}{rgb}{0.122, 0.467, 0.706}

%% file: results/main-results-table.tex
\begin{table*}[p]

\begin{subtable}{\textwidth}
\resizebox{\textwidth}{!}{
    \begin{tabular}{rl|lllllllllll|c}
    \toprule
    & & \textbf{Ar} & \textbf{Bn} & \textbf{En} & \textbf{Fi} & \textbf{Id} & \textbf{Ja} & \textbf{Ko} & \textbf{Ru} & \textbf{Sw} & \textbf{Te} & \textbf{Th} & \textbf{Avg} \\
    \midrule
    (1) & BM25 (default) & 
        0.368 & 0.418 & 0.140 & 0.284 & 0.376 & 0.211 & 0.285 & 0.313 & 0.389 & 0.343 & 0.401 & 0.321 \\
    (2) & BM25 (tuned) &
        0.367 & 0.413 & 0.151 & 0.288 & 0.382 & 0.217 & 0.281 & 0.329 & 0.396 & 0.424 & 0.417 & 0.333 \\
    \midrule
    (3) & mDPR (NQ pFT) & 
        0.291 & 0.291 & 0.291 & 0.206 & 0.271 & 0.213 & 0.235 & 0.283 & 0.189 & 0.111 & 0.172 & 0.226 \\ 
    (4) & mDPR (MS pFT) & 
        0.444 & 0.383 & 0.315 & 0.306 & 0.378 & 0.314 & 0.297 & 0.337 & 0.369 & 0.363 & 0.282 & 0.344 \\
    \midrule
    (5) & mDPR (MS pFT + in-lang FT)  &  0.691 & 0.651 & 0.489 & 0.551 & 0.562 & 0.488 & 0.453 & 0.485 & 0.640 & 0.876 & 0.619 & 0.591 \\
    (6) & mDPR (MS pFT + all FT) &  0.695 & 0.623 & 0.492 & 0.560 & 0.579 & 0.501 & 0.487 & 0.517 & 0.644 & 0.891 & 0.617 & 0.600 \\
    (7) & mDPR (MS pFT + in-script FT) &  -- & -- & 0.473 & 0.555 & 0.563 & -- & -- & -- & 0.635 & -- & -- &   -- \\
    (8) & mDPR (MS pFT + out-script FT) &  -- & -- & 0.476	& 0.563	& 0.565	& -- & -- & -- & 0.644	& -- & -- \\ 
    \midrule
    (9) & mDPR (in-lang FT) & 0.678 & 0.638 & 0.418 & 0.516 & 0.544 & 0.447 & 0.383 & 0.448  & 0.580 & 0.860 & 0.597  &  0.555 \\
    (10) & mDPR (all FT) & {0.695} & 0.659 & 0.476 & 0.550 & 0.565 & 0.496 & 0.453 & 0.515 & 0.633 & 0.891 & 0.607 & 0.594 \\
    (11) & mDPR (in-script FT) &  -- & -- & 0.444 & 0.535 & 0.560 & -- & -- & -- & 0.622 & -- & --  & --\\ 
    (12) & mDPR (out-script FT) &  -- & --	& 0.457	& 0.543	& 0.573	& -- & -- & --	& 0.624	& -- & -- \\	
    \midrule
    (13) & BM25 + row (6) & 
    0.714	& 0.702	& 0.520	& 0.590	& 0.634	& 0.558	& 0.523	& 0.590	& 0.623	& 0.845	& 0.697 & 0.636 \\
\midrule
\midrule
(a) & mono-ling DPR (in-lang FT) & 0.678 & -- & 0.426 & 0.573 & 0.545 & -- & 0.476 & -- & -- & -- & -- & --  \\
(b) & mono-ling DPR (in-script FT) & -- & -- & 0.412 & 0.540 & 0.488 & -- & -- & -- & -- & -- & -- & --  \\
(c) & mono-ling DPR (out-script FT) & 0.682 & -- & 0.426 & 0.522 & 0.540 & -- & 0.454 & -- & -- & -- & -- & --  \\
(d) & mono-ling DPR (all FT) & 0.682 & -- & 0.448 & 0.540 & 0.533 & -- & 0.454 & -- & -- & -- & -- & --     \\
\midrule
(e) & English DPR (in-lang FT) & 0.578 & 0.261 & 0.426 & 0.385 & 0.396 & 0.084 & 0.011 & 0.291 & 0.447 & 0.001 & 0.007 & 0.262    \\
(f) & English DPR (MS pFT + in-lang FT) & 0.592 & 0.318 & 0.497 & 0.423 & 0.439 & 0.218 & 0.182 & 0.298 & 0.499 & 0.001 & 0.030 & 0.318     \\
(g) & AfriBERTa DPR (in-lang FT) & 0.442 & 0.186 & 0.236 & 0.321 & 0.355 & 0.220 & 0.140 & 0.094 & 0.465 & 0.548 & 0.263 & 0.297     \\
\midrule
(h) & BM25 + row (f) & 0.628 & 0.480 & 0.501 & 0.480 & 0.510 & 0.333 & 0.299 & 0.440 & 0.535 & 0.423 & 0.424 & 0.459 \\ 
    \bottomrule
    \end{tabular}
}
\caption{MRR@100 on test set}
\label{tab:results_mrr}
\end{subtable}

\vspace{0.5em}

\begin{subtable}{\textwidth}
\resizebox{\textwidth}{!}{
    \begin{tabular}{rl|lllllllllll|c}
    \toprule
    & & \textbf{Ar} & \textbf{Bn} & \textbf{En} & \textbf{Fi} & \textbf{Id} & \textbf{Ja} & \textbf{Ko} & \textbf{Ru} & \textbf{Sw} & \textbf{Te} & \textbf{Th} & \textbf{Avg} \\
    \midrule
    (1) & BM25 (default) & 
        0.793 & 0.869 & 0.537 & 0.719 & 0.843 & 0.645 & 0.619 & 0.648 & 0.764 & 0.758 & 0.853 & 0.732 \\
    (2) & BM25 (tuned) & 
        0.800 & 0.874 & 0.551 & 0.725 & 0.846 & 0.656 & 0.797 & 0.660 & 0.764 & 0.813 & 0.853 & 0.758 \\
    \midrule
    (3) & mDPR (NQ pFT) & 
        0.650 & 0.779 & 0.678 & 0.568 & 0.685 & 0.584 & 0.533 & 0.647 & 0.528 & 0.366 & 0.515 & 0.594 \\
    (4) & mDPR (MS pFT) & 
        0.799 & 0.820 & 0.758 & 0.693 & 0.758 & 0.738 & 0.645 & 0.728 & 0.686 & 0.797 & 0.648 & 0.734 \\
    \midrule
    (5) & mDPR (MS pFT + in-lang FT) & 0.891 & 0.914 & 0.837 & 0.852 & 0.863 & 0.828 & 0.799 & 0.819 & 0.890 & 0.967 & 0.886 & 0.868 \\ 
    (6) & mDPR (MS pFT + all FT) & 0.900 & 0.955 & 0.841 & 0.856 & 0.860 & 0.813 & 0.785 & 0.843 & 0.876 & 0.966 & 0.883 & 0.871 \\ 
    (7) & mDPR (MS pFT + in-script FT) & 0.847 & 0.865 & 0.837 & 0.849 & 0.874 & -- & 0.740 & -- & 0.876 & 0.876 & 0.629  &  -- \\ 
    (8) & mDPR (MS pFT + out-script FT) &  -- & --	& 0.824	& 0.857	& 0.869	& -- & -- & -- & 0.873	& -- & -- \\ 
    \midrule
    (9) &mDPR (in-lang FT)  &  0.889 & 0.896 & 0.797 & 0.837 & 0.873 & 0.791 & 0.749 & 0.800 & 0.864 & 0.954 & 0.866 & 0.847 \\ 
    (10) & mDPR (all FT) & 0.894 & 0.937 & 0.839 & 0.846 & 0.867 & 0.811 & 0.771 & 0.819 & 0.893 & 0.969 & 0.866 & 0.865 \\ 
    (11) & mDPR (in-script FT)  &  -- & -- & 0.812 & 0.842 & 0.855 & -- & --- & -- & 0.877 & -- & --  &  -- \\ 
    (12) & mDPR (out-script FT) & -- & -- & 0.817 & 0.543 & 0.862 & -- & -- & --	& 0.876	& -- & --  \\
    \midrule
    (13) & BM25 + row (6) & 0.932 & 0.946	& 0.857	& 0.909	& 0.948	& 0.883	& 0.853	& 0.898	& 0.903	& 0.982	& 0.946 & 0.916 \\
\midrule
\midrule
(a) & mono-ling DPR (in-lang FT) & 0.894 & -- & 0.805 & 0.893 & 0.888 & -- & 0.820 & -- & -- & -- & -- & --  \\
(b) & mono-ling DPR (in-script FT) & -- & -- & 0.793 & 0.881 & 0.869 & -- & -- & -- & -- & -- & -- & --  \\
(c) & mono-ling DPR (out-script FT) & 0.890 & -- & 0.801 & 0.865 & 0.877 & -- & 0.802 & -- & -- & -- & -- & --  \\
(d) & mono-ling DPR (all FT) & 0.890 & -- & 0.814 & 0.867 & 0.890 & -- & 0.802 & -- & -- & -- & -- & --     \\
\midrule
(e) & English DPR (in-lang FT) & 0.807 & 0.581 & 0.805 & 0.712 & 0.740 & 0.297 & 0.021 & 0.689 & 0.790 & 0.017 & 0.035 & 0.499    \\
(f) & English DPR (MS pFT + in-lang FT) & 0.816 & 0.676 & 0.879 & 0.753 & 0.788 & 0.509 & 0.423 & 0.676 & 0.827 & 0.016 & 0.095 & 0.587     \\
(g) & AfriBERTa DPR (in-lang FT) & 0.738 & 0.451 & 0.565 & 0.658 & 0.692 & 0.633 & 0.416 & 0.293 & 0.808 & 0.872 & 0.657 & 0.617     \\
\midrule
(h) & BM25 + row (f) & 0.874 & 0.946 & 0.867 & 0.830 & 0.908 & 0.723 & 0.646 & 0.795 & 0.872 & 0.813 & 0.851 & 0.829 \\
    \bottomrule
    \end{tabular}
}
\caption{Recall@100 on test set}
\label{tab:results_recall1k}
\end{subtable}

\caption{
The effectiveness of different training conditions across all 11 languages in \mrtydi (and an overall average).
}
\label{table:main}
\end{table*}

%% file: results/lang2lang-table.tex
\begin{table*}[h]
\centering

\begin{subtable}[]{\columnwidth} 
\resizebox{\columnwidth}{!}{%
\begin{tabular}{lccccccccccc}
\hline
 & \textbf{Ar} & \textbf{Bn} & \textbf{En} & \textbf{Fi} & \textbf{In} & \textbf{Ja} & \textbf{Ko} & \textbf{Ru} & \textbf{Sw} & \textbf{Te} & \textbf{Th} \\ \hline
Ar & \cellcolor{table4left_0_0}{0.629} & \cellcolor{table4left_0_1}{0.560} & \cellcolor{table4left_0_2}{0.373} & \cellcolor{table4left_0_3}{0.455} & \cellcolor{table4left_0_4}{0.483} & \cellcolor{table4left_0_5}{0.367} & \cellcolor{table4left_0_6}{0.384} & \cellcolor{table4left_0_7}{0.451} & \cellcolor{table4left_0_8}{0.460} & \cellcolor{table4left_0_9}{0.506} & \cellcolor{table4left_0_10}{0.387}\\
Bn & \cellcolor{table4left_1_0}{0.448} & \cellcolor{table4left_1_1}{0.638} & \cellcolor{table4left_1_2}{0.309} & \cellcolor{table4left_1_3}{0.372} & \cellcolor{table4left_1_4}{0.383} & \cellcolor{table4left_1_5}{0.314} & \cellcolor{table4left_1_6}{0.326} & \cellcolor{table4left_1_7}{0.325} & \cellcolor{table4left_1_8}{0.433} & \cellcolor{table4left_1_9}{0.532} & \cellcolor{table4left_1_10}{0.341}\\
En & \cellcolor{table4left_2_0}{0.439} & \cellcolor{table4left_2_1}{0.389} & \cellcolor{table4left_2_2}{0.436} & \cellcolor{table4left_2_3}{0.381} & \cellcolor{table4left_2_4}{0.402} & \cellcolor{table4left_2_5}{0.296} & \cellcolor{table4left_2_6}{0.314} & \cellcolor{table4left_2_7}{0.363} & \cellcolor{table4left_2_8}{0.397} & \cellcolor{table4left_2_9}{0.289} & \cellcolor{table4left_2_10}{0.218}\\
Fi & \cellcolor{table4left_3_0}{0.524} & \cellcolor{table4left_3_1}{0.436} & \cellcolor{table4left_3_2}{0.369} & \cellcolor{table4left_3_3}{0.484} & \cellcolor{table4left_3_4}{0.471} & \cellcolor{table4left_3_5}{0.325} & \cellcolor{table4left_3_6}{0.358} & \cellcolor{table4left_3_7}{0.420} & \cellcolor{table4left_3_8}{0.470} & \cellcolor{table4left_3_9}{0.619} & \cellcolor{table4left_3_10}{0.295}\\
In & \cellcolor{table4left_4_0}{0.486} & \cellcolor{table4left_4_1}{0.490} & \cellcolor{table4left_4_2}{0.375} & \cellcolor{table4left_4_3}{0.415} & \cellcolor{table4left_4_4}{0.524} & \cellcolor{table4left_4_5}{0.326} & \cellcolor{table4left_4_6}{0.353} & \cellcolor{table4left_4_7}{0.372} & \cellcolor{table4left_4_8}{0.442} & \cellcolor{table4left_4_9}{0.366} & \cellcolor{table4left_4_10}{0.221}\\
Ja & \cellcolor{table4left_5_0}{0.480} & \cellcolor{table4left_5_1}{0.573} & \cellcolor{table4left_5_2}{0.328} & \cellcolor{table4left_5_3}{0.391} & \cellcolor{table4left_5_4}{0.395} & \cellcolor{table4left_5_5}{0.471} & \cellcolor{table4left_5_6}{0.382} & \cellcolor{table4left_5_7}{0.401} & \cellcolor{table4left_5_8}{0.441} & \cellcolor{table4left_5_9}{0.615} & \cellcolor{table4left_5_10}{0.345}\\
Ko & \cellcolor{table4left_6_0}{0.421} & \cellcolor{table4left_6_1}{0.464} & \cellcolor{table4left_6_2}{0.315} & \cellcolor{table4left_6_3}{0.373} & \cellcolor{table4left_6_4}{0.377} & \cellcolor{table4left_6_5}{0.309} & \cellcolor{table4left_6_6}{0.383} & \cellcolor{table4left_6_7}{0.330} & \cellcolor{table4left_6_8}{0.421} & \cellcolor{table4left_6_9}{0.300} & \cellcolor{table4left_6_10}{0.181}\\
Ru & \cellcolor{table4left_7_0}{0.476} & \cellcolor{table4left_7_1}{0.524} & \cellcolor{table4left_7_2}{0.337} & \cellcolor{table4left_7_3}{0.410} & \cellcolor{table4left_7_4}{0.400} & \cellcolor{table4left_7_5}{0.364} & \cellcolor{table4left_7_6}{0.311} & \cellcolor{table4left_7_7}{0.419} & \cellcolor{table4left_7_8}{0.424} & \cellcolor{table4left_7_9}{0.531} & \cellcolor{table4left_7_10}{0.243}\\
Sw & \cellcolor{table4left_8_0}{0.422} & \cellcolor{table4left_8_1}{0.397} & \cellcolor{table4left_8_2}{0.251} & \cellcolor{table4left_8_3}{0.324} & \cellcolor{table4left_8_4}{0.367} & \cellcolor{table4left_8_5}{0.258} & \cellcolor{table4left_8_6}{0.304} & \cellcolor{table4left_8_7}{0.317} & \cellcolor{table4left_8_8}{0.580} & \cellcolor{table4left_8_9}{0.276} & \cellcolor{table4left_8_10}{0.228}\\
Te & \cellcolor{table4left_9_0}{0.512} & \cellcolor{table4left_9_1}{0.587} & \cellcolor{table4left_9_2}{0.331} & \cellcolor{table4left_9_3}{0.399} & \cellcolor{table4left_9_4}{0.404} & \cellcolor{table4left_9_5}{0.398} & \cellcolor{table4left_9_6}{0.368} & \cellcolor{table4left_9_7}{0.380} & \cellcolor{table4left_9_8}{0.421} & \cellcolor{table4left_9_9}{0.861} & \cellcolor{table4left_9_10}{0.422}\\
Th & \cellcolor{table4left_10_0}{0.505} & \cellcolor{table4left_10_1}{0.554} & \cellcolor{table4left_10_2}{0.319} & \cellcolor{table4left_10_3}{0.389} & \cellcolor{table4left_10_4}{0.410} & \cellcolor{table4left_10_5}{0.361} & \cellcolor{table4left_10_6}{0.323} & \cellcolor{table4left_10_7}{0.416} & \cellcolor{table4left_10_8}{0.411} & \cellcolor{table4left_10_9}{0.648} & \cellcolor{table4left_10_10}{0.588}\\
\bottomrule
\end{tabular}
}
\vspace{0.1cm}
\caption{mDPR initialized from raw mBERT}
\label{tab:lang2lang-mbert}
\end{subtable}
\hfill
\begin{subtable}[]{\columnwidth}
\resizebox{0.99\columnwidth}{!}{%
\begin{tabular}{lccccccccccc}

\hline
 & \textbf{Ar} & \textbf{Bn} & \textbf{En} & \textbf{Fi} & \textbf{In} & \textbf{Ja} & \textbf{Ko} & \textbf{Ru} & \textbf{Sw} & \textbf{Te} & \textbf{Th} \\ \hline
Ar & \cellcolor{table4right_0_0}{0.663} & \cellcolor{table4right_0_1}{0.608} & \cellcolor{table4right_0_2}{0.450} & \cellcolor{table4right_0_3}{0.450} & \cellcolor{table4right_0_4}{0.508} & \cellcolor{table4right_0_5}{0.432} & \cellcolor{table4right_0_6}{0.392} & \cellcolor{table4right_0_7}{0.509} & \cellcolor{table4right_0_8}{0.425} & \cellcolor{table4right_0_9}{0.613} & \cellcolor{table4right_0_10}{0.380}\\
Bn & \cellcolor{table4right_1_0}{0.549} & \cellcolor{table4right_1_1}{0.645} & \cellcolor{table4right_1_2}{0.432} & \cellcolor{table4right_1_3}{0.427} & \cellcolor{table4right_1_4}{0.468} & \cellcolor{table4right_1_5}{0.432} & \cellcolor{table4right_1_6}{0.401} & \cellcolor{table4right_1_7}{0.435} & \cellcolor{table4right_1_8}{0.438} & \cellcolor{table4right_1_9}{0.644} & \cellcolor{table4right_1_10}{0.417}\\
En & \cellcolor{table4right_2_0}{0.469} & \cellcolor{table4right_2_1}{0.412} & \cellcolor{table4right_2_2}{0.480} & \cellcolor{table4right_2_3}{0.292} & \cellcolor{table4right_2_4}{0.319} & \cellcolor{table4right_2_5}{0.368} & \cellcolor{table4right_2_6}{0.344} & \cellcolor{table4right_2_7}{0.376} & \cellcolor{table4right_2_8}{0.302} & \cellcolor{table4right_2_9}{0.283} & \cellcolor{table4right_2_10}{0.245}\\
Fi & \cellcolor{table4right_3_0}{0.542} & \cellcolor{table4right_3_1}{0.501} & \cellcolor{table4right_3_2}{0.472} & \cellcolor{table4right_3_3}{0.524} & \cellcolor{table4right_3_4}{0.495} & \cellcolor{table4right_3_5}{0.396} & \cellcolor{table4right_3_6}{0.409} & \cellcolor{table4right_3_7}{0.477} & \cellcolor{table4right_3_8}{0.441} & \cellcolor{table4right_3_9}{0.463} & \cellcolor{table4right_3_10}{0.355}\\
In & \cellcolor{table4right_4_0}{0.551} & \cellcolor{table4right_4_1}{0.538} & \cellcolor{table4right_4_2}{0.432} & \cellcolor{table4right_4_3}{0.441} & \cellcolor{table4right_4_4}{0.561} & \cellcolor{table4right_4_5}{0.385} & \cellcolor{table4right_4_6}{0.396} & \cellcolor{table4right_4_7}{0.443} & \cellcolor{table4right_4_8}{0.430} & \cellcolor{table4right_4_9}{0.496} & \cellcolor{table4right_4_10}{0.349}\\
Ja & \cellcolor{table4right_5_0}{0.531} & \cellcolor{table4right_5_1}{0.500} & \cellcolor{table4right_5_2}{0.374} & \cellcolor{table4right_5_3}{0.367} & \cellcolor{table4right_5_4}{0.405} & \cellcolor{table4right_5_5}{0.483} & \cellcolor{table4right_5_6}{0.396} & \cellcolor{table4right_5_7}{0.453} & \cellcolor{table4right_5_8}{0.385} & \cellcolor{table4right_5_9}{0.429} & \cellcolor{table4right_5_10}{0.375}\\
Ko & \cellcolor{table4right_6_0}{0.515} & \cellcolor{table4right_6_1}{0.450} & \cellcolor{table4right_6_2}{0.430} & \cellcolor{table4right_6_3}{0.379} & \cellcolor{table4right_6_4}{0.428} & \cellcolor{table4right_6_5}{0.426} & \cellcolor{table4right_6_6}{0.453} & \cellcolor{table4right_6_7}{0.499} & \cellcolor{table4right_6_8}{0.413} & \cellcolor{table4right_6_9}{0.404} & \cellcolor{table4right_6_10}{0.336}\\
Ru & \cellcolor{table4right_7_0}{0.570} & \cellcolor{table4right_7_1}{0.521} & \cellcolor{table4right_7_2}{0.448} & \cellcolor{table4right_7_3}{0.437} & \cellcolor{table4right_7_4}{0.488} & \cellcolor{table4right_7_5}{0.427} & \cellcolor{table4right_7_6}{0.400} & \cellcolor{table4right_7_7}{0.499} & \cellcolor{table4right_7_8}{0.421} & \cellcolor{table4right_7_9}{0.510} & \cellcolor{table4right_7_10}{0.331}\\
Sw & \cellcolor{table4right_8_0}{0.498} & \cellcolor{table4right_8_1}{0.470} & \cellcolor{table4right_8_2}{0.388} & \cellcolor{table4right_8_3}{0.409} & \cellcolor{table4right_8_4}{0.440} & \cellcolor{table4right_8_5}{0.365} & \cellcolor{table4right_8_6}{0.353} & \cellcolor{table4right_8_7}{0.428} & \cellcolor{table4right_8_8}{0.635} & \cellcolor{table4right_8_9}{0.494} & \cellcolor{table4right_8_10}{0.331}\\
Te & \cellcolor{table4right_9_0}{0.575} & \cellcolor{table4right_9_1}{0.583} & \cellcolor{table4right_9_2}{0.426} & \cellcolor{table4right_9_3}{0.390} & \cellcolor{table4right_9_4}{0.433} & \cellcolor{table4right_9_5}{0.449} & \cellcolor{table4right_9_6}{0.413} & \cellcolor{table4right_9_7}{0.480} & \cellcolor{table4right_9_8}{0.432} & \cellcolor{table4right_9_9}{0.867} & \cellcolor{table4right_9_10}{0.436}\\
Th & \cellcolor{table4right_10_0}{0.587} & \cellcolor{table4right_10_1}{0.530} & \cellcolor{table4right_10_2}{0.446} & \cellcolor{table4right_10_3}{0.410} & \cellcolor{table4right_10_4}{0.490} & \cellcolor{table4right_10_5}{0.427} & \cellcolor{table4right_10_6}{0.411} & \cellcolor{table4right_10_7}{0.480} & \cellcolor{table4right_10_8}{0.428} & \cellcolor{table4right_10_9}{0.605} & \cellcolor{table4right_10_10}{0.611}\\

\bottomrule
\end{tabular}
}
\vspace{0.1cm}
\caption{mDPR pre--fine-tuned with MS MARCO}
\label{tab:lang2lang-ms}
\end{subtable}

\caption{
Matrix experiment training mDPR on the language denoted in the row, evaluating on the language in the column, reporting MRR@100.
On the left, fine-tuning is performed on the raw mBERT backbone directly.
On the right, fine-tuning is performed after pFT with MS MARCO passage. For Bn, Ko and Sw, we use all available training data; for the remaining languages, the amount of training data is sampled to 3,300 training queries.
The values are highlighted \textit{column-wise}:\ the color gradient ranges from \textcolor{blue}{\textbf{blue}} (high values) to \textcolor{orange}{\textbf{orange}} (low values). 
}
\label{table:mdpr-matrix}
\vspace{-0.3cm}
\end{table*}

%% file: mdpr.bbl

\begin{thebibliography}{36}


\ifx \showCODEN    \undefined \def \showCODEN     #1{\unskip}     \fi
\ifx \showDOI      \undefined \def \showDOI       #1{#1}\fi
\ifx \showISBNx    \undefined \def \showISBNx     #1{\unskip}     \fi
\ifx \showISBNxiii \undefined \def \showISBNxiii  #1{\unskip}     \fi
\ifx \showISSN     \undefined \def \showISSN      #1{\unskip}     \fi
\ifx \showLCCN     \undefined \def \showLCCN      #1{\unskip}     \fi
\ifx \shownote     \undefined \def \shownote      #1{#1}          \fi
\ifx \showarticletitle \undefined \def \showarticletitle #1{#1}   \fi
\ifx \showURL      \undefined \def \showURL       {\relax}        \fi
\providecommand\bibfield[2]{#2}
\providecommand\bibinfo[2]{#2}
\providecommand\natexlab[1]{#1}
\providecommand\showeprint[2][]{arXiv:#2}

\bibitem[\protect\citeauthoryear{Armengol-Estap{\'e}, Carrino,
  Rodriguez-Penagos, de~Gibert~Bonet, Armentano-Oller, Gonzalez-Agirre, Melero,
  and Villegas}{Armengol-Estap{\'e} et~al\mbox{.}}{2021}]%
        {armengol-estape-etal-2021-multilingual}
\bibfield{author}{\bibinfo{person}{Jordi Armengol-Estap{\'e}},
  \bibinfo{person}{Casimiro~Pio Carrino}, \bibinfo{person}{Carlos
  Rodriguez-Penagos}, \bibinfo{person}{Ona de Gibert~Bonet},
  \bibinfo{person}{Carme Armentano-Oller}, \bibinfo{person}{Aitor
  Gonzalez-Agirre}, \bibinfo{person}{Maite Melero}, {and}
  \bibinfo{person}{Marta Villegas}.} \bibinfo{year}{2021}\natexlab{}.
\newblock \showarticletitle{Are Multilingual Models the Best Choice for
  Moderately Under-resourced Languages? {A} Comprehensive Assessment for
  {C}atalan}. In \bibinfo{booktitle}{\emph{Findings of the Association for
  Computational Linguistics: ACL-IJCNLP 2021}}. \bibinfo{address}{Online},
  \bibinfo{pages}{4933--4946}.
\newblock


\bibitem[\protect\citeauthoryear{Artetxe, Ruder, and Yogatama}{Artetxe
  et~al\mbox{.}}{2020}]%
        {artetxe-etal-2020-cross}
\bibfield{author}{\bibinfo{person}{Mikel Artetxe}, \bibinfo{person}{Sebastian
  Ruder}, {and} \bibinfo{person}{Dani Yogatama}.}
  \bibinfo{year}{2020}\natexlab{}.
\newblock \showarticletitle{On the Cross-lingual Transferability of Monolingual
  Representations}. In \bibinfo{booktitle}{\emph{Proceedings of the 58th Annual
  Meeting of the Association for Computational Linguistics}}.
  \bibinfo{address}{Online}, \bibinfo{pages}{4623--4637}.
\newblock


\bibitem[\protect\citeauthoryear{Asai, Kasai, Clark, Lee, Choi, and
  Hajishirzi}{Asai et~al\mbox{.}}{2021a}]%
        {Asai:2010.11856:2021}
\bibfield{author}{\bibinfo{person}{Akari Asai}, \bibinfo{person}{Jungo Kasai},
  \bibinfo{person}{Jonathan~H. Clark}, \bibinfo{person}{Kenton Lee},
  \bibinfo{person}{Eunsol Choi}, {and} \bibinfo{person}{Hannaneh Hajishirzi}.}
  \bibinfo{year}{2021}\natexlab{a}.
\newblock \showarticletitle{{XOR} {QA}: Cross-lingual Open-Retrieval Question
  Answering}.
\newblock \bibinfo{journal}{\emph{arXiv:2010.11856}} (\bibinfo{year}{2021}).
\newblock


\bibitem[\protect\citeauthoryear{Asai, Yu, Kasai, and Hajishirzi}{Asai
  et~al\mbox{.}}{2021b}]%
        {asai-2021-one}
\bibfield{author}{\bibinfo{person}{Akari Asai}, \bibinfo{person}{Xinyan Yu},
  \bibinfo{person}{Jungo Kasai}, {and} \bibinfo{person}{Hannaneh Hajishirzi}.}
  \bibinfo{year}{2021}\natexlab{b}.
\newblock \showarticletitle{One Question Answering Model for Many Languages
  with Cross-lingual Dense Passage Retrieval}.
\newblock \bibinfo{journal}{\emph{arXiv:2107.11976}} (\bibinfo{year}{2021}).
\newblock


\bibitem[\protect\citeauthoryear{Bonifacio, Jeronymo, Abonizio, Campiotti,
  Fadaee, Lotufo, and Nogueira}{Bonifacio et~al\mbox{.}}{2022}]%
        {Bonifacio:2108.13897:2022}
\bibfield{author}{\bibinfo{person}{Luiz~Henrique Bonifacio},
  \bibinfo{person}{Vitor Jeronymo}, \bibinfo{person}{Hugo~Queiroz Abonizio},
  \bibinfo{person}{Israel Campiotti}, \bibinfo{person}{Marzieh Fadaee},
  \bibinfo{person}{Roberto Lotufo}, {and} \bibinfo{person}{Rodrigo Nogueira}.}
  \bibinfo{year}{2022}\natexlab{}.
\newblock \showarticletitle{{mMARCO}: A Multilingual Version of the MS MARCO
  Passage Ranking Dataset}.
\newblock \bibinfo{journal}{\emph{arXiv:2108.13897}} (\bibinfo{year}{2022}).
\newblock


\bibitem[\protect\citeauthoryear{Boytsov, Novak, Malkov, and Nyberg}{Boytsov
  et~al\mbox{.}}{2016}]%
        {Boytsov_etal_CIKM2016}
\bibfield{author}{\bibinfo{person}{Leonid Boytsov}, \bibinfo{person}{David
  Novak}, \bibinfo{person}{Yury Malkov}, {and} \bibinfo{person}{Eric Nyberg}.}
  \bibinfo{year}{2016}\natexlab{}.
\newblock \showarticletitle{Off the Beaten Path: Let's Replace Term-Based
  Retrieval with {k-NN} Search}. In \bibinfo{booktitle}{\emph{Proceedings of
  25th International Conference on Information and Knowledge Management (CIKM
  2016)}}. \bibinfo{address}{Indianapolis, Indiana},
  \bibinfo{pages}{1099--1108}.
\newblock


\bibitem[\protect\citeauthoryear{Clark, Choi, Collins, Garrette, Kwiatkowski,
  Nikolaev, and Palomaki}{Clark et~al\mbox{.}}{2020}]%
        {tydi}
\bibfield{author}{\bibinfo{person}{Jonathan~H. Clark}, \bibinfo{person}{Eunsol
  Choi}, \bibinfo{person}{Michael Collins}, \bibinfo{person}{Dan Garrette},
  \bibinfo{person}{Tom Kwiatkowski}, \bibinfo{person}{Vitaly Nikolaev}, {and}
  \bibinfo{person}{Jennimaria Palomaki}.} \bibinfo{year}{2020}\natexlab{}.
\newblock \showarticletitle{{T}y{D}i {QA}: A Benchmark for Information-Seeking
  Question Answering in Typologically Diverse Languages}.
\newblock \bibinfo{journal}{\emph{Transactions of the Association for
  Computational Linguistics}}  \bibinfo{volume}{8} (\bibinfo{year}{2020}),
  \bibinfo{pages}{454--470}.
\newblock


\bibitem[\protect\citeauthoryear{Conneau, Wu, Li, Zettlemoyer, and
  Stoyanov}{Conneau et~al\mbox{.}}{2020}]%
        {conneau-etal-2020-emerging}
\bibfield{author}{\bibinfo{person}{Alexis Conneau}, \bibinfo{person}{Shijie
  Wu}, \bibinfo{person}{Haoran Li}, \bibinfo{person}{Luke Zettlemoyer}, {and}
  \bibinfo{person}{Veselin Stoyanov}.} \bibinfo{year}{2020}\natexlab{}.
\newblock \showarticletitle{Emerging Cross-lingual Structure in Pretrained
  Language Models}. In \bibinfo{booktitle}{\emph{Proceedings of the 58th Annual
  Meeting of the Association for Computational Linguistics}}.
  \bibinfo{address}{Online}, \bibinfo{pages}{6022--6034}.
\newblock


\bibitem[\protect\citeauthoryear{Devlin, Chang, Lee, and Toutanova}{Devlin
  et~al\mbox{.}}{2019}]%
        {devlin-etal-2019-bert}
\bibfield{author}{\bibinfo{person}{Jacob Devlin}, \bibinfo{person}{Ming-Wei
  Chang}, \bibinfo{person}{Kenton Lee}, {and} \bibinfo{person}{Kristina
  Toutanova}.} \bibinfo{year}{2019}\natexlab{}.
\newblock \showarticletitle{{BERT}: Pre-training of Deep Bidirectional
  Transformers for Language Understanding}. In
  \bibinfo{booktitle}{\emph{Proceedings of the 2019 Conference of the North
  {A}merican Chapter of the Association for Computational Linguistics: Human
  Language Technologies, Volume 1 (Long and Short Papers)}}.
  \bibinfo{address}{Minneapolis, Minnesota}, \bibinfo{pages}{4171--4186}.
\newblock


\bibitem[\protect\citeauthoryear{Dufter and Sch{\"u}tze}{Dufter and
  Sch{\"u}tze}{2020}]%
        {dufter-schutze-2020-identifying}
\bibfield{author}{\bibinfo{person}{Philipp Dufter} {and}
  \bibinfo{person}{Hinrich Sch{\"u}tze}.} \bibinfo{year}{2020}\natexlab{}.
\newblock \showarticletitle{Identifying Elements Essential for {BERT}{'}s
  Multilinguality}. In \bibinfo{booktitle}{\emph{Proceedings of the 2020
  Conference on Empirical Methods in Natural Language Processing (EMNLP)}}.
  \bibinfo{address}{Online}, \bibinfo{pages}{4423--4437}.
\newblock


\bibitem[\protect\citeauthoryear{Galuščáková, Oard, and
  Nair}{Galuščáková et~al\mbox{.}}{2021}]%
        {clir-survey}
\bibfield{author}{\bibinfo{person}{Petra Galuščáková},
  \bibinfo{person}{Douglas~W. Oard}, {and} \bibinfo{person}{Suraj Nair}.}
  \bibinfo{year}{2021}\natexlab{}.
\newblock \showarticletitle{Cross-language Information Retrieval}.
\newblock \bibinfo{journal}{\emph{arXiv:2111.05988}} (\bibinfo{year}{2021}).
\newblock


\bibitem[\protect\citeauthoryear{Garg, Vu, and Moschitti}{Garg
  et~al\mbox{.}}{2020}]%
        {Garg_etal_AAAI2020}
\bibfield{author}{\bibinfo{person}{Siddhant Garg}, \bibinfo{person}{Thuy Vu},
  {and} \bibinfo{person}{Alessandro Moschitti}.}
  \bibinfo{year}{2020}\natexlab{}.
\newblock \showarticletitle{{TANDA}: Transfer and Adapt Pre-Trained Transformer
  Models for Answer Sentence Selection}. In
  \bibinfo{booktitle}{\emph{Proceedings of the Thirty-Fourth AAAI Conference on
  Artificial Intelligence (AAAI-20)}}. \bibinfo{address}{New York, New York},
  \bibinfo{pages}{7780--7788}.
\newblock


\bibitem[\protect\citeauthoryear{Hofst\"{a}tter, Lin, Yang, Lin, and
  Hanbury}{Hofst\"{a}tter et~al\mbox{.}}{2021}]%
        {tasb}
\bibfield{author}{\bibinfo{person}{Sebastian Hofst\"{a}tter},
  \bibinfo{person}{Sheng-Chieh Lin}, \bibinfo{person}{Jheng-Hong Yang},
  \bibinfo{person}{Jimmy Lin}, {and} \bibinfo{person}{Allan Hanbury}.}
  \bibinfo{year}{2021}\natexlab{}.
\newblock \showarticletitle{Efficiently Teaching an Effective Dense Retriever
  with Balanced Topic Aware Sampling}. In \bibinfo{booktitle}{\emph{Proceedings
  of the 44th International ACM SIGIR Conference on Research and Development in
  Information Retrieval}}. \bibinfo{pages}{113–122}.
\newblock
\showISBNx{9781450380379}


\bibitem[\protect\citeauthoryear{Johnson, Douze, and {J\'{e}gou}}{Johnson
  et~al\mbox{.}}{2021}]%
        {Johnson_etal_2021}
\bibfield{author}{\bibinfo{person}{Jeff Johnson}, \bibinfo{person}{Matthijs
  Douze}, {and} \bibinfo{person}{{Herv\'{e}} {J\'{e}gou}}.}
  \bibinfo{year}{2021}\natexlab{}.
\newblock \showarticletitle{Billion-scale similarity search with {GPUs}}.
\newblock \bibinfo{journal}{\emph{IEEE Transactions on Big Data}}
  \bibinfo{volume}{7}, \bibinfo{number}{3} (\bibinfo{year}{2021}),
  \bibinfo{pages}{535--547}.
\newblock


\bibitem[\protect\citeauthoryear{K, Wang, Mayhew, and Roth}{K
  et~al\mbox{.}}{2019}]%
        {wang2019cross}
\bibfield{author}{\bibinfo{person}{Karthikeyan K}, \bibinfo{person}{Zihan
  Wang}, \bibinfo{person}{Stephen Mayhew}, {and} \bibinfo{person}{Dan Roth}.}
  \bibinfo{year}{2019}\natexlab{}.
\newblock \showarticletitle{Cross-Lingual Ability of Multilingual {BERT:} An
  Empirical Study}.
\newblock \bibinfo{journal}{\emph{arXiv:1912.07840}} (\bibinfo{year}{2019}).
\newblock


\bibitem[\protect\citeauthoryear{Karpukhin, Oguz, Min, Lewis, Wu, Edunov, Chen,
  and Yih}{Karpukhin et~al\mbox{.}}{2020}]%
        {dpr}
\bibfield{author}{\bibinfo{person}{Vladimir Karpukhin}, \bibinfo{person}{Barlas
  Oguz}, \bibinfo{person}{Sewon Min}, \bibinfo{person}{Patrick Lewis},
  \bibinfo{person}{Ledell Wu}, \bibinfo{person}{Sergey Edunov},
  \bibinfo{person}{Danqi Chen}, {and} \bibinfo{person}{Wen-tau Yih}.}
  \bibinfo{year}{2020}\natexlab{}.
\newblock \showarticletitle{Dense Passage Retrieval for Open-Domain Question
  Answering}. In \bibinfo{booktitle}{\emph{Proceedings of the 2020 Conference
  on Empirical Methods in Natural Language Processing (EMNLP)}}.
  \bibinfo{address}{Online}, \bibinfo{pages}{6769--6781}.
\newblock


\bibitem[\protect\citeauthoryear{Lauscher, Ravishankar, Vuli{\'c}, and
  Glava{\v{s}}}{Lauscher et~al\mbox{.}}{2020}]%
        {lauscher-etal-2020-zero}
\bibfield{author}{\bibinfo{person}{Anne Lauscher}, \bibinfo{person}{Vinit
  Ravishankar}, \bibinfo{person}{Ivan Vuli{\'c}}, {and} \bibinfo{person}{Goran
  Glava{\v{s}}}.} \bibinfo{year}{2020}\natexlab{}.
\newblock \showarticletitle{From Zero to Hero: {O}n the Limitations of
  Zero-Shot Language Transfer with Multilingual {T}ransformers}. In
  \bibinfo{booktitle}{\emph{Proceedings of the 2020 Conference on Empirical
  Methods in Natural Language Processing (EMNLP)}}. \bibinfo{address}{Online},
  \bibinfo{pages}{4483--4499}.
\newblock


\bibitem[\protect\citeauthoryear{Lin, Yang, and Lin}{Lin et~al\mbox{.}}{2021}]%
        {tct_colbert}
\bibfield{author}{\bibinfo{person}{Sheng-Chieh Lin},
  \bibinfo{person}{Jheng-Hong Yang}, {and} \bibinfo{person}{Jimmy Lin}.}
  \bibinfo{year}{2021}\natexlab{}.
\newblock \showarticletitle{In-Batch Negatives for Knowledge Distillation with
  Tightly-Coupled Teachers for Dense Retrieval}. In
  \bibinfo{booktitle}{\emph{Proceedings of the 6th Workshop on Representation
  Learning for NLP (RepL4NLP-2021)}}. \bibinfo{address}{Online},
  \bibinfo{pages}{163--173}.
\newblock


\bibitem[\protect\citeauthoryear{Litschko, Vulic, Ponzetto, and
  Glavas}{Litschko et~al\mbox{.}}{2021}]%
        {litschko-2021-evaluating-multilingual}
\bibfield{author}{\bibinfo{person}{Robert Litschko}, \bibinfo{person}{Ivan
  Vulic}, \bibinfo{person}{Simone~Paolo Ponzetto}, {and} \bibinfo{person}{Goran
  Glavas}.} \bibinfo{year}{2021}\natexlab{}.
\newblock \showarticletitle{Evaluating Multilingual Text Encoders for
  Unsupervised Cross-Lingual Retrieval}.
\newblock \bibinfo{journal}{\emph{arXiv:2101.08370}} (\bibinfo{year}{2021}).
\newblock


\bibitem[\protect\citeauthoryear{MacAvaney, Soldaini, and Goharian}{MacAvaney
  et~al\mbox{.}}{2020}]%
        {MacAvaney2020TeachingAN}
\bibfield{author}{\bibinfo{person}{Sean MacAvaney}, \bibinfo{person}{Luca
  Soldaini}, {and} \bibinfo{person}{Nazli Goharian}.}
  \bibinfo{year}{2020}\natexlab{}.
\newblock \showarticletitle{Teaching a New Dog Old Tricks: Resurrecting
  Multilingual Retrieval Using Zero-Shot Learning}. In
  \bibinfo{booktitle}{\emph{Proceedings of the 42nd European Conference on
  Information Retrieval, Part II (ECIR 2020)}}. \bibinfo{pages}{246--254}.
\newblock


\bibitem[\protect\citeauthoryear{Martin, Muller, Ortiz~Su{\'a}rez, Dupont,
  Romary, de~la Clergerie, Seddah, and Sagot}{Martin et~al\mbox{.}}{2020}]%
        {martin-etal-2020-camembert}
\bibfield{author}{\bibinfo{person}{Louis Martin}, \bibinfo{person}{Benjamin
  Muller}, \bibinfo{person}{Pedro~Javier Ortiz~Su{\'a}rez},
  \bibinfo{person}{Yoann Dupont}, \bibinfo{person}{Laurent Romary},
  \bibinfo{person}{{\'E}ric de~la Clergerie}, \bibinfo{person}{Djam{\'e}
  Seddah}, {and} \bibinfo{person}{Beno{\^\i}t Sagot}.}
  \bibinfo{year}{2020}\natexlab{}.
\newblock \showarticletitle{{C}amem{BERT}: a Tasty {F}rench Language Model}. In
  \bibinfo{booktitle}{\emph{Proceedings of the 58th Annual Meeting of the
  Association for Computational Linguistics}}. \bibinfo{address}{Online},
  \bibinfo{pages}{7203--7219}.
\newblock


\bibitem[\protect\citeauthoryear{Nie}{Nie}{2010}]%
        {Nie_2010}
\bibfield{author}{\bibinfo{person}{Jian-Yun Nie}.}
  \bibinfo{year}{2010}\natexlab{}.
\newblock \bibinfo{booktitle}{\emph{Cross-Language Information Retrieval}}.
\newblock \bibinfo{publisher}{Morgan \& Claypool Publishers}.
\newblock


\bibitem[\protect\citeauthoryear{Ogueji, Zhu, and Lin}{Ogueji
  et~al\mbox{.}}{2021}]%
        {ogueji-etal-2021-small}
\bibfield{author}{\bibinfo{person}{Kelechi Ogueji}, \bibinfo{person}{Yuxin
  Zhu}, {and} \bibinfo{person}{Jimmy Lin}.} \bibinfo{year}{2021}\natexlab{}.
\newblock \showarticletitle{Small Data? No Problem! Exploring the Viability of
  Pretrained Multilingual Language Models for Low-resourced Languages}. In
  \bibinfo{booktitle}{\emph{Proceedings of the 1st Workshop on Multilingual
  Representation Learning}}. \bibinfo{address}{Punta Cana, Dominican Republic},
  \bibinfo{pages}{116--126}.
\newblock


\bibitem[\protect\citeauthoryear{Pires, Schlinger, and Garrette}{Pires
  et~al\mbox{.}}{2019}]%
        {pires-etal-2019-multilingual}
\bibfield{author}{\bibinfo{person}{Telmo Pires}, \bibinfo{person}{Eva
  Schlinger}, {and} \bibinfo{person}{Dan Garrette}.}
  \bibinfo{year}{2019}\natexlab{}.
\newblock \showarticletitle{How Multilingual is Multilingual {BERT}?}. In
  \bibinfo{booktitle}{\emph{Proceedings of the 57th Annual Meeting of the
  Association for Computational Linguistics}}. \bibinfo{address}{Florence,
  Italy}, \bibinfo{pages}{4996--5001}.
\newblock


\bibitem[\protect\citeauthoryear{R{\"o}nnqvist, Kanerva, Salakoski, and
  Ginter}{R{\"o}nnqvist et~al\mbox{.}}{2019}]%
        {ronnqvist-etal-2019-multilingual}
\bibfield{author}{\bibinfo{person}{Samuel R{\"o}nnqvist},
  \bibinfo{person}{Jenna Kanerva}, \bibinfo{person}{Tapio Salakoski}, {and}
  \bibinfo{person}{Filip Ginter}.} \bibinfo{year}{2019}\natexlab{}.
\newblock \showarticletitle{Is Multilingual {BERT} Fluent in Language
  Generation?}. In \bibinfo{booktitle}{\emph{Proceedings of the First NLPL
  Workshop on Deep Learning for Natural Language Processing}}.
  \bibinfo{address}{Turku, Finland}, \bibinfo{pages}{29--36}.
\newblock


\bibitem[\protect\citeauthoryear{Shi, Bai, and Lin}{Shi et~al\mbox{.}}{2020}]%
        {Shi2020CrossLingualTO}
\bibfield{author}{\bibinfo{person}{Peng Shi}, \bibinfo{person}{He Bai}, {and}
  \bibinfo{person}{Jimmy Lin}.} \bibinfo{year}{2020}\natexlab{}.
\newblock \showarticletitle{Cross-Lingual Training of Neural Models for
  Document Ranking}. In \bibinfo{booktitle}{\emph{Findings of the Association
  for Computational Linguistics: EMNLP 2020}}. \bibinfo{pages}{2768--2773}.
\newblock


\bibitem[\protect\citeauthoryear{Shi, Zhang, Bai, and Lin}{Shi
  et~al\mbox{.}}{2021}]%
        {shi-etal-2021-cross}
\bibfield{author}{\bibinfo{person}{Peng Shi}, \bibinfo{person}{Rui Zhang},
  \bibinfo{person}{He Bai}, {and} \bibinfo{person}{Jimmy Lin}.}
  \bibinfo{year}{2021}\natexlab{}.
\newblock \showarticletitle{Cross-Lingual Training of Dense Retrievers for
  Document Retrieval}. In \bibinfo{booktitle}{\emph{Proceedings of the 1st
  Workshop on Multilingual Representation Learning}}. \bibinfo{address}{Punta
  Cana, Dominican Republic}, \bibinfo{pages}{251--253}.
\newblock


\bibitem[\protect\citeauthoryear{Sun and Duh}{Sun and Duh}{2020}]%
        {sun-duh-2020-clirmatrix}
\bibfield{author}{\bibinfo{person}{Shuo Sun} {and} \bibinfo{person}{Kevin
  Duh}.} \bibinfo{year}{2020}\natexlab{}.
\newblock \showarticletitle{{CLIRM}atrix: A massively large collection of
  bilingual and multilingual datasets for Cross-Lingual Information Retrieval}.
  In \bibinfo{booktitle}{\emph{Proceedings of the 2020 Conference on Empirical
  Methods in Natural Language Processing (EMNLP)}}. \bibinfo{address}{Online},
  \bibinfo{pages}{4160--4170}.
\newblock


\bibitem[\protect\citeauthoryear{Virtanen, Kanerva, Ilo, Luoma, Luotolahti,
  Salakoski, Ginter, and Pyysalo}{Virtanen et~al\mbox{.}}{2019}]%
        {virtanen2019multilingual}
\bibfield{author}{\bibinfo{person}{Antti Virtanen}, \bibinfo{person}{Jenna
  Kanerva}, \bibinfo{person}{Rami Ilo}, \bibinfo{person}{Jouni Luoma},
  \bibinfo{person}{Juhani Luotolahti}, \bibinfo{person}{Tapio Salakoski},
  \bibinfo{person}{Filip Ginter}, {and} \bibinfo{person}{Sampo Pyysalo}.}
  \bibinfo{year}{2019}\natexlab{}.
\newblock \showarticletitle{Multilingual is not enough: BERT for Finnish}.
\newblock \bibinfo{journal}{\emph{arXiv:1912.07076}} (\bibinfo{year}{2019}).
\newblock


\bibitem[\protect\citeauthoryear{Wolf, Debut, Sanh, Chaumond, Delangue, Moi,
  Cistac, Rault, Louf, Funtowicz, Davison, Shleifer, von Platen, Ma, Jernite,
  Plu, Xu, Le~Scao, Gugger, Drame, Lhoest, and Rush}{Wolf
  et~al\mbox{.}}{2020}]%
        {wolf-etal-2020-transformers}
\bibfield{author}{\bibinfo{person}{Thomas Wolf}, \bibinfo{person}{Lysandre
  Debut}, \bibinfo{person}{Victor Sanh}, \bibinfo{person}{Julien Chaumond},
  \bibinfo{person}{Clement Delangue}, \bibinfo{person}{Anthony Moi},
  \bibinfo{person}{Pierric Cistac}, \bibinfo{person}{Tim Rault},
  \bibinfo{person}{Remi Louf}, \bibinfo{person}{Morgan Funtowicz},
  \bibinfo{person}{Joe Davison}, \bibinfo{person}{Sam Shleifer},
  \bibinfo{person}{Patrick von Platen}, \bibinfo{person}{Clara Ma},
  \bibinfo{person}{Yacine Jernite}, \bibinfo{person}{Julien Plu},
  \bibinfo{person}{Canwen Xu}, \bibinfo{person}{Teven Le~Scao},
  \bibinfo{person}{Sylvain Gugger}, \bibinfo{person}{Mariama Drame},
  \bibinfo{person}{Quentin Lhoest}, {and} \bibinfo{person}{Alexander Rush}.}
  \bibinfo{year}{2020}\natexlab{}.
\newblock \showarticletitle{Transformers: State-of-the-Art Natural Language
  Processing}. In \bibinfo{booktitle}{\emph{Proceedings of the 2020 Conference
  on Empirical Methods in Natural Language Processing: System Demonstrations}}.
  \bibinfo{address}{Online}, \bibinfo{pages}{38--45}.
\newblock


\bibitem[\protect\citeauthoryear{Wu and Dredze}{Wu and Dredze}{2019}]%
        {wu-dredze-2019-beto}
\bibfield{author}{\bibinfo{person}{Shijie Wu} {and} \bibinfo{person}{Mark
  Dredze}.} \bibinfo{year}{2019}\natexlab{}.
\newblock \showarticletitle{Beto, Bentz, Becas: The Surprising Cross-Lingual
  Effectiveness of {BERT}}. In \bibinfo{booktitle}{\emph{Proceedings of the
  2019 Conference on Empirical Methods in Natural Language Processing and the
  9th International Joint Conference on Natural Language Processing
  (EMNLP-IJCNLP)}}. \bibinfo{address}{Hong Kong, China},
  \bibinfo{pages}{833--844}.
\newblock


\bibitem[\protect\citeauthoryear{Xie, Yang, Tan, Xiong, Yuan, Huai, Li, and
  Lin}{Xie et~al\mbox{.}}{2020}]%
        {Xie_etal_WWW2020}
\bibfield{author}{\bibinfo{person}{Yuqing Xie}, \bibinfo{person}{Wei Yang},
  \bibinfo{person}{Luchen Tan}, \bibinfo{person}{Kun Xiong},
  \bibinfo{person}{Nicholas~Jing Yuan}, \bibinfo{person}{Baoxing Huai},
  \bibinfo{person}{Ming Li}, {and} \bibinfo{person}{Jimmy Lin}.}
  \bibinfo{year}{2020}\natexlab{}.
\newblock \showarticletitle{Distant Supervision for Multi-Stage Fine-Tuning in
  Retrieval-Based Question Answering}. In \bibinfo{booktitle}{\emph{Proceedings
  of The Web Conference 2020 (WWW '20)}}. \bibinfo{pages}{2934--2940}.
\newblock


\bibitem[\protect\citeauthoryear{Xiong, Xiong, Li, Tang, Liu, Bennett, Ahmed,
  and Overwijk}{Xiong et~al\mbox{.}}{2021}]%
        {xiong21ance}
\bibfield{author}{\bibinfo{person}{Lee Xiong}, \bibinfo{person}{Chenyan Xiong},
  \bibinfo{person}{Ye Li}, \bibinfo{person}{Kwok{-}Fung Tang},
  \bibinfo{person}{Jialin Liu}, \bibinfo{person}{Paul~N. Bennett},
  \bibinfo{person}{Junaid Ahmed}, {and} \bibinfo{person}{Arnold Overwijk}.}
  \bibinfo{year}{2021}\natexlab{}.
\newblock \showarticletitle{Approximate Nearest Neighbor Negative Contrastive
  Learning for Dense Text Retrieval}. In \bibinfo{booktitle}{\emph{Proceedings
  of the 9th International Conference on Learning Representations (ICLR
  2021)}}.
\newblock


\bibitem[\protect\citeauthoryear{Zhang, Westerfield, Shim, Bingham, Fabbri, Hu,
  Verma, and Radev}{Zhang et~al\mbox{.}}{2019}]%
        {zhang-etal-2019-improving-low}
\bibfield{author}{\bibinfo{person}{Rui Zhang}, \bibinfo{person}{Caitlin
  Westerfield}, \bibinfo{person}{Sungrok Shim}, \bibinfo{person}{Garrett
  Bingham}, \bibinfo{person}{Alexander Fabbri}, \bibinfo{person}{William Hu},
  \bibinfo{person}{Neha Verma}, {and} \bibinfo{person}{Dragomir Radev}.}
  \bibinfo{year}{2019}\natexlab{}.
\newblock \showarticletitle{Improving Low-Resource Cross-lingual Document
  Retrieval by Reranking with Deep Bilingual Representations}. In
  \bibinfo{booktitle}{\emph{Proceedings of the 57th Annual Meeting of the
  Association for Computational Linguistics}}. \bibinfo{address}{Florence,
  Italy}, \bibinfo{pages}{3173--3179}.
\newblock


\bibitem[\protect\citeauthoryear{Zhang, Ma, Shi, and Lin}{Zhang
  et~al\mbox{.}}{2021a}]%
        {mrtydi}
\bibfield{author}{\bibinfo{person}{Xinyu Zhang}, \bibinfo{person}{Xueguang Ma},
  \bibinfo{person}{Peng Shi}, {and} \bibinfo{person}{Jimmy Lin}.}
  \bibinfo{year}{2021}\natexlab{a}.
\newblock \showarticletitle{Mr. {T}y{D}i: A Multi-lingual Benchmark for Dense
  Retrieval}. In \bibinfo{booktitle}{\emph{Proceedings of the 1st Workshop on
  Multilingual Representation Learning}}. \bibinfo{address}{Punta Cana,
  Dominican Republic}, \bibinfo{pages}{127--137}.
\newblock


\bibitem[\protect\citeauthoryear{Zhang, Yates, and Lin}{Zhang
  et~al\mbox{.}}{2021b}]%
        {ZhangXinyu_etal_ECIR2021}
\bibfield{author}{\bibinfo{person}{Xinyu Zhang}, \bibinfo{person}{Andrew
  Yates}, {and} \bibinfo{person}{Jimmy Lin}.} \bibinfo{year}{2021}\natexlab{b}.
\newblock \showarticletitle{Comparing Score Aggregation Approaches for Document
  Retrieval with Pretrained Transformers}. In
  \bibinfo{booktitle}{\emph{Proceedings of the 43rd European Conference on
  Information Retrieval (ECIR 2021), Part II}}. \bibinfo{pages}{150--163}.
\newblock


\end{thebibliography}
